\documentclass[10pt,a4paper,ptm]{article}
\usepackage[utf8]{inputenc}
\usepackage[english]{babel}
\usepackage{amsmath}
\usepackage{amsfonts}
\usepackage{amssymb}
\usepackage{graphicx}
\usepackage{xcolor,tikz,datetime,multirow,lscape,multicol,float,soul}
\usepackage[outline]{contour}
\contourlength{.2pt}
\usepackage[round]{natbib}
\usepackage[affil-it]{authblk}
\usepackage[breaklinks=true,bookmarks=true,colorlinks=true,linkcolor=blue,citecolor=blue]{hyperref}

\usepackage{soul,xcolor,multirow,ifthen,float,morefloats}
\usepackage{geometry}
\newlength\halflineskip
\newlength\affilskip
\usepackage{fancyhdr}
\usepackage[font=small,labelfont=sc]{caption}
\usepackage{titlesec}
\titleformat{\section}
  {\bf\large}
  {\thesection}{1em}{\vspace*{-.2cm}}
\usepackage{etoolbox}
\AtBeginEnvironment{tabular}{\small}

\usepackage{array}
\newcommand{\bibdir}{./}
\newcommand{\figdir}{./figure/}

\newcommand{\mmt}[1]{{\color{blue!}#1}}

\newcommand{\mut}[1]{{\color{red!}#1}}

\newcommand{\cbr}[1]{\textcolor{black}{\contour{red}{#1}}}
\newcommand{\cbg}[1]{\textcolor{black}{\contour{green!60!blue}{#1}}}

\newcommand{\putpic}[3]{\put(#1){\includegraphics[trim=0cm 0cm 0cm 0cm, clip, width=#2\textwidth]{\figdir #3}}}
\newif\ifref
\reffalse
\newif\ifall
\alltrue
\newif\ifone
\newif\iftwo
\newif\ifthree
\newif\iffour
\newif\iffive
\newif\ifsix
\newif\ifseven
\newif\ifeight
\newif\ifnine
\newif\iften
\newif\ifeleven
\newif\iftwelve
\newif\ifthirteen
\newif\iffourteen
\newif\iffifteen
\newif\ifsixteen
\newif\ifseventeen
\newif\ifeighteen
\newif\ifnineteen
\newif\iftwenty
\newif\iftwentyone
\onefalse
\twofalse
\threefalse
\fourfalse
\fivefalse
\sixfalse
\sevenfalse
\eightfalse
\ninefalse
\tenfalse
\elevenfalse
\twelvefalse
\thirteenfalse
\fourteenfalse
\fifteenfalse
\sixteenfalse
\seventeenfalse
\eighteenfalse
\nineteenfalse
\twentyfalse
\twentyonefalse
\ifall
\onetrue
\twotrue
\threetrue
\fourtrue
\fivetrue
\sixtrue
\seventrue
\eighttrue
\ninetrue
\tentrue
\eleventrue
\twelvetrue
\thirteentrue
\fourteentrue
\fifteentrue
\sixteentrue
\seventeentrue
\eighteentrue
\nineteentrue
\twentytrue
\twentyonetrue
\fi
\twentyonetrue
\thirteentrue

\newsavebox{\astrutbox}
\sbox{\astrutbox}{\rule[-5pt]{0pt}{20pt}}

\title{Direct numerical simulation of the oscillatory flow around a sphere resting on a rough bottom}

\author[1]{{\bf Marco Mazzuoli}}
\author[1]{{\bf Paolo Blondeaux}}
\author[2]{{\bf Julian Simeonov}}
\author[2]{{\bf Joseph Calantoni}}
\affil[1]{{\small Department of Civil, Chemical and Environmental Engineering (DICCA), University of Genoa, 
Via Montallegro 1, 16145 Genova, Italy}}
\affil[2]{{\small Marine Geosciences Division - Naval Research Laboratory - Stennis Space Center 
Mississipi, U.S.A.}}

\date{May, 2017}

\begin{document}

\maketitle

\vspace{.5cm}

\begin{abstract}
The oscillatory flow around a spherical object lying on a rough bottom is investigated 
by means of direct numerical simulations of continuity and Navier-Stokes equations. 
The rough bottom is simulated by a layer/multiple layers of spherical particles, 
the size of which is much smaller that the size of the object. 
The period and amplitude of the velocity oscillations of the free stream are chosen
to mimic the flow at the bottom of sea waves and the size of the small spherical 
particles falls in the range of coarse sand/very fine gravel. 
Even though the computational costs allow only the simulation of moderate values of the
Reynolds number characterizing the bottom boundary layer, the results show that 
the coherent vortex structures, shed by the spherical object, can break-up and generate 
turbulence, if the Reynolds number of the object is sufficiently large. 
The knowledge of the velocity field allows the dynamics of the large scale coherent 
vortices shed by the object to be determined and turbulence characteristics to be evaluated. 
Moreover, the forces and torques acting on both the large spherical object and the 
small particles, simulating sediment grains, can be determined and analysed, thus 
laying the groundwork for the investigation of sediment dynamics and scour developments.
\end{abstract}


\section{Introduction}
\label{introd}

An object lying on the sea bed causes a local acceleration of the flow 
field and a local increase of the bottom shear stress. 
It follows that the sediments surrounding the object might be swept away 
from it, causing a local lowering of the bed profile,
even when the flow far from the object is not strong enough to move the sediment. 
This phenomenon is observed at different spatial scales, which range from that of a 
small pebble lying on a sandy bottom to that of the foundation of a large coastal structure. 
Then, the scour which develops around the object has different consequences as, 
for example, the self-buring of the object (e.g. self-burial of a pipeline) 
and its possible instability (e.g. the instability of a wind mill). 
The self-buring of the object is also quite important for mines or unexploded 
ordnance (UXO).
For example, nowdays, because of increasing human activities in shallow waters 
(e.g. navigation, fisheries, sand extraction), the buried mines and UXO of World 
Wars I and II are a threat to public safety and remediation in many coastal areas 
is becoming a priority.

A key role in the mechanics of sediment transport and in the dynamics of the 
scour around the object is played by the dynamics of the vortices which are 
generated by the interaction of the external flow with the object. 
Indeed, these vortices tend to pick-up the sediment grains from the bottom, 
to make them rolling and sliding along the bottom surface or even to carry them 
into suspension, when the vertical velocity they induce is larger than the fall 
velocity of the sediment particles. 

The determination of the threshold conditions, above which the sediments start 
to move or are carried into suspension, is a complex problem. 
Indeed, to quantify the bottom erodibility, it is necessary to know both the 
mechanical properties of the sediments (e.g. grain size and sediment density) 
and the dynamics of the large vortex structures which are present close to 
the bottom and can induce, locally, large values of the hydrodynamic forces 
acting on sediment particles. 

In coastal environments, the phenomenon is made more complex by the oscillatory character of the flow induced by the propagation 
of sea waves, which makes the vortex structures shed by the object during a half cycle to interact with the object and the vortices 
shed during the previous half cycle. This nonlinear interaction might give rise to a possible chaotic flow which might appear through different scenarios, e.g. 
Feigenbaum scenario \citep{blondeaux1991} and quasi-periodicity and phase-locking scenario \citep{vittori1993}.
 
Moreover, for relatively small objects, the vortices shed by the object might interact with the small eddies shed by the sediment 
grains and the eddies generated by the transition process from the laminar to the turbulent regime in the bottom boundary layer.
On the other hand, for relatively large objects, the vortices shed by the object do not interact directly with the small eddies shed by the 
sediment grains and the turbulent eddies, even though the latter certainly affect the dynamics of the former.

Therefore, to obtain an accurate and reliable description of the phenomenon, it is necessary to consider the simultaneous presence
of i) the vortex structures shed by the object, ii) the small vortices shed by the sediment grains and iii)
the possible presence of turbulent eddies. The turbulent eddies appear when the Reynolds number of the flow is large enough to trigger the 
transition process from the laminar to the turbulent regime either in the bottom boundary layer or in the free shear layers released
by the object invested by the oscillatory flow.

For an oscillatory boundary layer over a smooth wall, both experimental measurements \citep[e.g.][]{Hino1976} and direct numerical 
simulations \citep[e.g.][]{verzicco1996,Vittori1998} indicate that turbulence appears explosively during 
the decelerating phases of the oscillatory cycle, when the Reynolds number $R_\delta$ is larger than a value ranging between $500$ and $600$. 
Hereinafter, the Reynolds number is defined with the amplitude $U^*_0$ of the velocity oscillations far from the bottom and the thickness
of the viscous bottom boundary layer $\delta^*=\sqrt{2 \nu^*/\omega^*}$, $\nu^*$ being the kinematic viscosity of the water
and $\omega^*$ the angular frequency of the velocity oscillations. However, close to the critical conditions, turbulence does not survive during the 
accelerating phases and the flow recovers a laminar 'like' behaviour ('intermittently turbulent regime'). 
Larger values of the Reynolds number cause turbulence to appear earlier and to pervade larger parts of the cycle till, at high 
Reynolds numbers, turbulence is present throughout the cycle. 
The direct numerical simulations of \citet{Costamagna2003} showed that the elementary process which generates turbulent eddies 
in an oscillatory flow is similar to that in a steady flow. 

More recently, \citet{Carstensen2010} observed the presence of turbulent spots 
during the transition process from the laminar to the turbulent regime in an
oscillatory boundary layer and showed that these isolated turbulent areas in 
an otherwise laminar boundary layer, cause violent oscillations of both the 
velocity and the shear stress. 
Moreover, \citet{Carstensen2010} observed that turbulent spots emerge from the 
breaking of low speed streaks. 
Later, the existence of turbulent spots in an oscillatory boundary layer was 
confirmed by the direct numerical simulations of \citet{Mazzuoli2011}. 
Since the numerical simulations give access to velocity and pressure fields 
in the three-dimensional space and time, the numerical results of \citet{Mazzuoli2011} 
supplemented the experimental measurements of \citet{Carstensen2010} and in 
particular allowed to determine the speed of the head and tail of the spots 
along with the speed of the lateral spreading of the spot.

As already pointed out, the studies summarized so far were carried out by considering a smooth bottom.
In natural environments, the sediment grains make the bottom to be rough and generate small vortices, 
which interact with the turbulent eddies. 
An experimental investigation of the oscillatory flow over macro-roughness elements
was made by \citet{Sleath1976}, who measured the velocity profile in an 
oscillatory boundary layer over spheres of large diameter, arranged in an hexagonal 
pattern. The measurements of \citet{Sleath1976} showed a complex turbulent flow 
field and suggested the existence of coherent vortex structures which are shed by the roughness 
elements at flow reversal and move away from the bottom. The oscillatory flow over a
similar rough bottom was investigated by \citet{Fornarelli2009} by means of
direct numerical simulations of continuity and Navier-Stokes equations. The roughness
consisted of semi-spheres regularly fixed on a plane wall in an hexagonal pattern.
The results of \citet{Fornarelli2009} show that the temporal development of
the velocity close to the spheres is characterized by two maxima. One maximum is correlated
to the maximum of the free stream velocity. A further peak in the velocity appears close
to the reversal of the external flow and is generated by the passage of the vortex structures
shed by the roughness elements, which move away from the bottom. 

More recently, the oscillatory flow over a layer of spherical grains has been simulated by 
\citet{Mazzuoli2016} for different values of the diameter of the spheres and different values of
the Reynolds number. Their results show that at least three flow regimes exist,
namely the laminar regime, the transitional turbulent regime and the hydrodynamically rough turbulent regime.
For relatively small values of the Reynolds number, the turbulent kinetic energy has negligible values and the flow
regime can be defined laminar. When the Reynolds number is increased, two different regimes are encountered depending
on the value of the sphere size. For small spheres, it is likely that the turbulent regime is due
to an intrinsically instability of the oscillatory boundary layer. Indeed, turbulent fluctuations are observed when
the Reynolds number is larger than a critical value similar to that of the Stokes boundary layer over a flat wall. 
On the other hand, for large spheres, turbulence is generated by the nonlinear interaction of the free shear layers 
shed by the sediment grains.  

There are many other results on the flow over a bottom of regular roughness elements. For example, let us mention 
that, recently, \citet{celik2014} have carried out pressure measurements on a spherical grain resting upon a
bed of identical grains. However, even though interesting results have been obtained by \citet{celik2014}, their
results as well as other results which are not summarized herein consider a steady forcing flow, the characteristics
of which are different from those of an oscillatory flow.

Much less is known on the dynamics of the three-dimensional vortex structures shed by an object lying on a flat wall and subject to an 
oscillatory flow. \citet{fischer2002} made direct numerical simulations of the oscillatory flow around a sphere laying
on a plane wall. The investigation of \citet{fischer2002} was inspired by the experiments described by \citet{rosenthal1986} 
and was aimed at providing more information on the lift and drag forces acting on a sediment grain at the bottom of sea waves.
Unlike the results of \citet{CherukMcLau1994}, \citet{cherukat1994}, \citet{asmolov1999a} and \citet{asmolov1999}, the numerical result of \citet{fischer2002}
are not restricted to relatively small values of the Reynolds number or to disparate diffusive, convective and oscillatory length scales.
However, attention was focused on the forces acting an the sphere and the velocity and vorticity field around the sphere as well as the shear stress
acting on the bottom were not analysed.

Let us mention also the recent studies of the flow and scour around pipelines and piles of \citet{fuhrman2014} and \citet{baykal2015l}
where the results of previous studies are also summarized. The investigations of \citet{fuhrman2014} and \citet{baykal2015l} were carried
out by solving the Reynolds averaged Navier-Stokes (RANS) equations and introducing a two-equation turbulence model. The introduction of the
Reynolds average has the advantage of allowing the simulation of flow fields characterized by high Reynolds numbers, however,
a RANS approach does not resolve explicitly the turbulent mixing and hence does not provide accurate results about the dynamics of the
coherent vortices shed by the objects and their interaction with the turbulent eddies and the vortices shed by the sediment grains.

The present paper describes the results of direct numerical simulations of the oscillatory 
flow over an idealized sea bottom made by small spherical particles, which simulate 
a coarse sand or a very fine gravel sediment, above which a much larger sphere is resting, which can be though to represent 
any object at rest (e.g. a small cobble or a small unexploded ordnance). 
The use  of direct numerical simulations allows us to evaluate quantities that are very
difficult to measure in a laboratory experiment (e.g. vorticity, dissipation and production
of turbulence, ...). Moreover, the results of direct numerical simulations allow a
detailed study of the vortex structures generated during the oscillatory cycle to be carried out, 
along with the investigation of the interaction of the vortices with the particles lying on the bottom.

The forcing flow is assumed to be oscillatory and generated by the
propagation of a surface wave. Hence, the frequency of the fluid oscillations is chosen to 
reproduce what happens at the bottom of a sea wave. The computational costs do not allow high Renyolds numbers to be simulated
but results for values of the Reynolds number large enough to trigger transition to turbulence are obtained and presented.

The structure of the rest of the paper is the following. In the next section, we formulate the problem and summarize the main steps of the numerical
procedure employed to determine the oscillatory flow around a spherical object resting on the sea bed. In section~3, we describe the results 
focusing the attention on the dynamics of the coherent vortex structures shed by the object and on their interaction with the bottom and the roughness
elements. Section~4 for is devoted to the conclusions and to a brief description of the future developments of the work.

\section{Formulation of the problem and numerical approach}

\begin{figure}
\begin{picture}(0,250)(0,0)
\ifone
\put(60,0){
\put(00,0){\includegraphics[trim=0cm 0cm 0cm 0cm, clip, width=.6\textwidth]{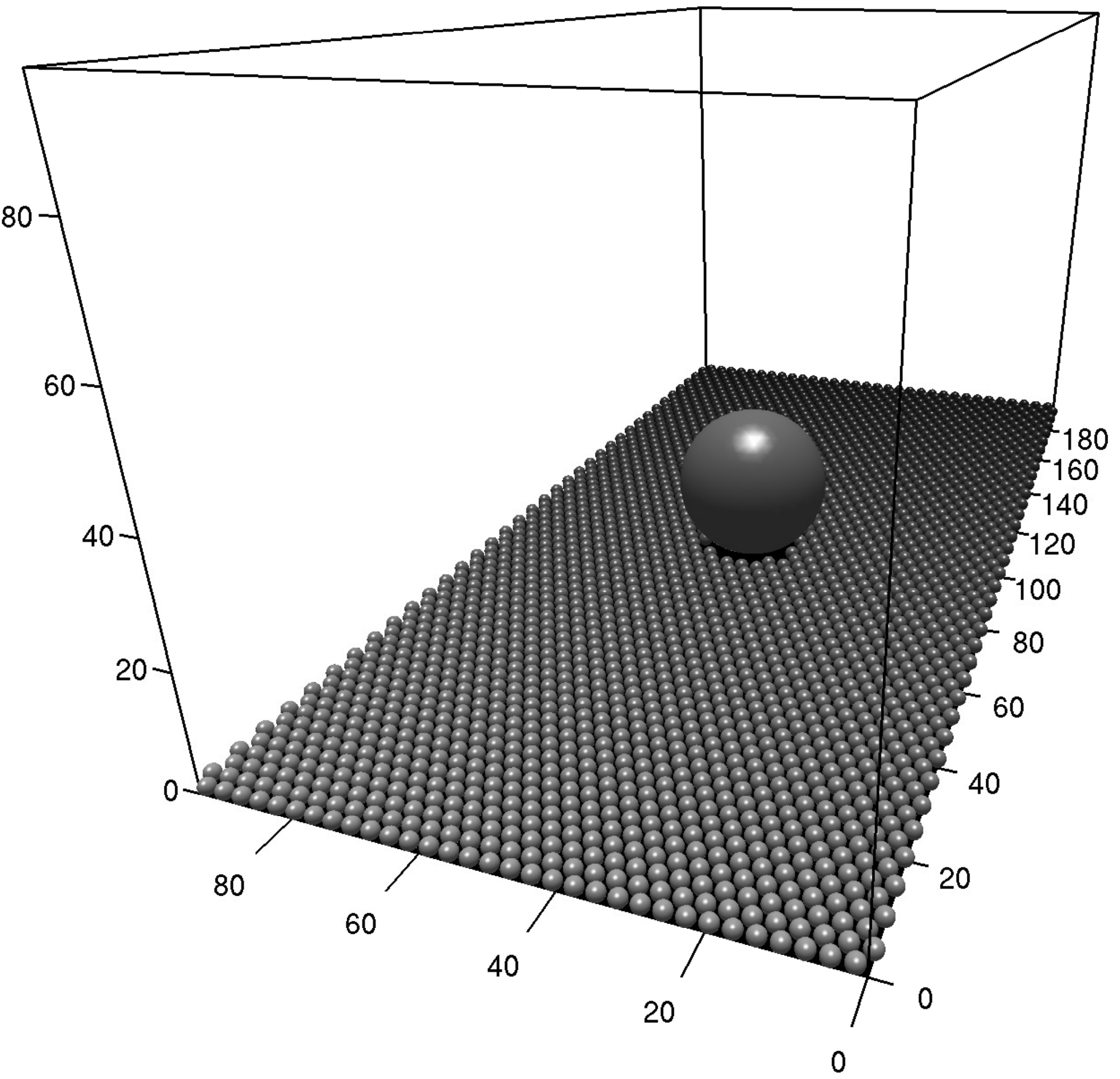}}
\put(220,80){$x_1$}
\put(85,10){$x_2$}
\put(-5,125){$x_3$}
}
\fi
\end{picture}
\caption{
Sketch of the problem (e.g. run with $R_\delta=56$, $D=28$, $d=2.6$ and $\epsilon=0.374$).
}
\label{fig1}
\end{figure}
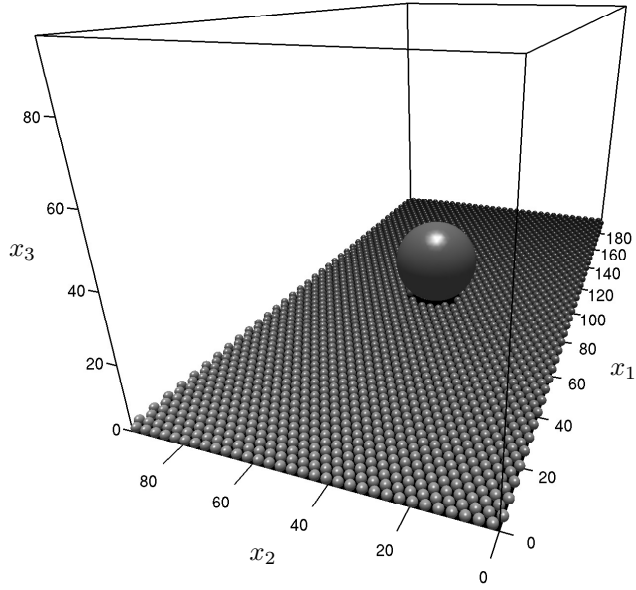
%
\begin{table}
	\begin{center}
		\begin{tabular}{l c c c c c c c c c c c}
		\hline
		\multirow{2}{*}{$R_\delta$} &
		\multirow{2}{*}{$D$} & 
		\multirow{2}{*}{$d$} & 
		\multirow{2}{1.4cm}{no. layers \\ of spheres} & 
		\multirow{2}{*}{$\epsilon$} & 
		\multirow{2}{*}{$L_{x_1}$} & 
		\multirow{2}{*}{$L_{x_2}$} &
		\multirow{2}{*}{$L_{x_3}$} &
		\multirow{2}{*}{$\Delta x_{fine}$} &
		\multirow{2}{*}{$\Delta x_{coarse}$} &
		\multirow{2}{1cm}{$\displaystyle\frac{D}{\Delta x_{fine}}$} &
		\multirow{2}{1.3cm}{no. small \\ spheres} \vspace*{.4cm}\\ \hline
		$47.87$ & $6.267$ &  $-$   & $-$ & $0.09777$ & $60.54$ & $60.54$ & $30.27$ & $0.059$ & $0.24$ & $106$ & $-$ \\
		$56.0$  & $28.0$  & $2.6$  & $1$ & $0.374$   & $191.3$ & $95.66$ & $95.66$ & $0.19$  & $0.75$ & $150$ & $2338$\\
		$112.1$ & $28.0$  & $2.6$  & $1$ & $0.374$   & $382.6$ & $95.66$ & $95.66$ & $0.19$  & $0.75$ & $150$ & $4740$\\
		$112.1$ & $28.0$  & $2.6$  & $5$ & $0.374$   & $382.6$ & $95.66$ & $95.66$ & $0.19$  & $1.50$ & $150$ & $24548$ \\
		$112.1$ & $11.2$  & $0.56$ & $3$ & $0.112$   & $172.2$ & $43.05$ & $43.05$ & $0.056$ & $0.45$ & $200$ & $56620$ \\
		\hline
		\end{tabular}
	\end{center}
	\caption{\small Summary of domain discretization and flow parameters for the present runs.}
	\label{tab1}
\end{table}

When a propagating surface wave of small amplitude is considered,
it is well known that, at the leading order of approximation and in a region the thickness of which
scales with the amplitude of the fluid oscillations, the flow close to the bottom 
can be studied by considering the flow generated close to a wall by an oscillating pressure gradient described by 
\begin{equation}
\label{pres}
\frac{\partial p^*}{\partial x^*_1} = - \rho^* U^*_0 \omega^*
\sin (\omega^*t^*); \ \ \ \ \frac{\partial p^*}{\partial x^*_2} =
0; \ \ \ \ \frac{\partial p^*}{\partial x^*_3} = 0
\end{equation}
where ($x_1^*, x_2^*,x_3^*$) is a Cartesian coordinate system with the $x^*_1$-axis pointing 
in the direction of wave propagation and the $x_3^*$-axis being vertical and pointing in the upward 
direction. 
In (\ref{pres}), $\rho^*$ is the density of the sea water, assumed to be constant, and $U^*_0$ 
and $\omega^*=2 \pi/T^*$ are the amplitude and the angular frequency of the fluid velocity 
oscillations induced by the surface wave close to the bottom but in the region
where the flow is irrotational and the fluid behaves like an inviscid fluid. 
Hereinafter, a star is used to denote dimensional quantities, while the same symbols 
without the star denote their dimensionless counterparts.

\begin{figure}
\begin{picture}(0,625)(0,0)
\iftwo
\put(60,500){
\putpic{0,0}{.65}{figure2a}
\put(-5,66){$x_3$}
\put(-2,115){$a)$}
}
\put(60,375){
\putpic{0,0}{.65}{figure2b}
\put(-5,66){$x_3$}
\put(-2,115){$b)$}
}
\put(60,250){
\putpic{0,0}{.65}{figure2c}
\put(-5,66){$x_3$}
\put(-2,115){$c)$}
}
\put(60,125){
\putpic{0,0}{.65}{figure2d}
\put(-5,66){$x_3$}
\put(-2,115){$d)$}
}
\put(60,0){
\putpic{0,0}{.65}{figure2e}
\put(-5,66){$x_3$}
\put(120,-5){$x_1$}
\put(-2,115){$e)$}
}
\fi
\end{picture}
\caption{
Spanwise component of the vorticity computed in the middle vertical plane crossing the sphere ($x_2 = 30$) for 
$R_\delta=47.87$, $D=6.267$, $\epsilon=\epsilon^*/\delta^*=0.09777$
at $t = \omega^* t^*= 3.66\pi$, panel (a); $t=3.80\pi$, panel (b); $t=3.96\pi$, panel (c); $t=4.44\pi$, panel (d); $t=4.50\pi$, panel (e).
The vorticity isolines are for $\omega^*_2 \delta^*/U^*_0 = \pm 0.2, \pm 0.5, \pm 1, \pm 2, \pm 3, \pm 4$ (solid lines = positive values, broken lines = negative values).
When comparing the present results with those by \citet{fischer2002}, the reader should consider that the values of the vorticity isolines in Fischer et al.'s paper
are unknown.}
\label{figA}
\end{figure}
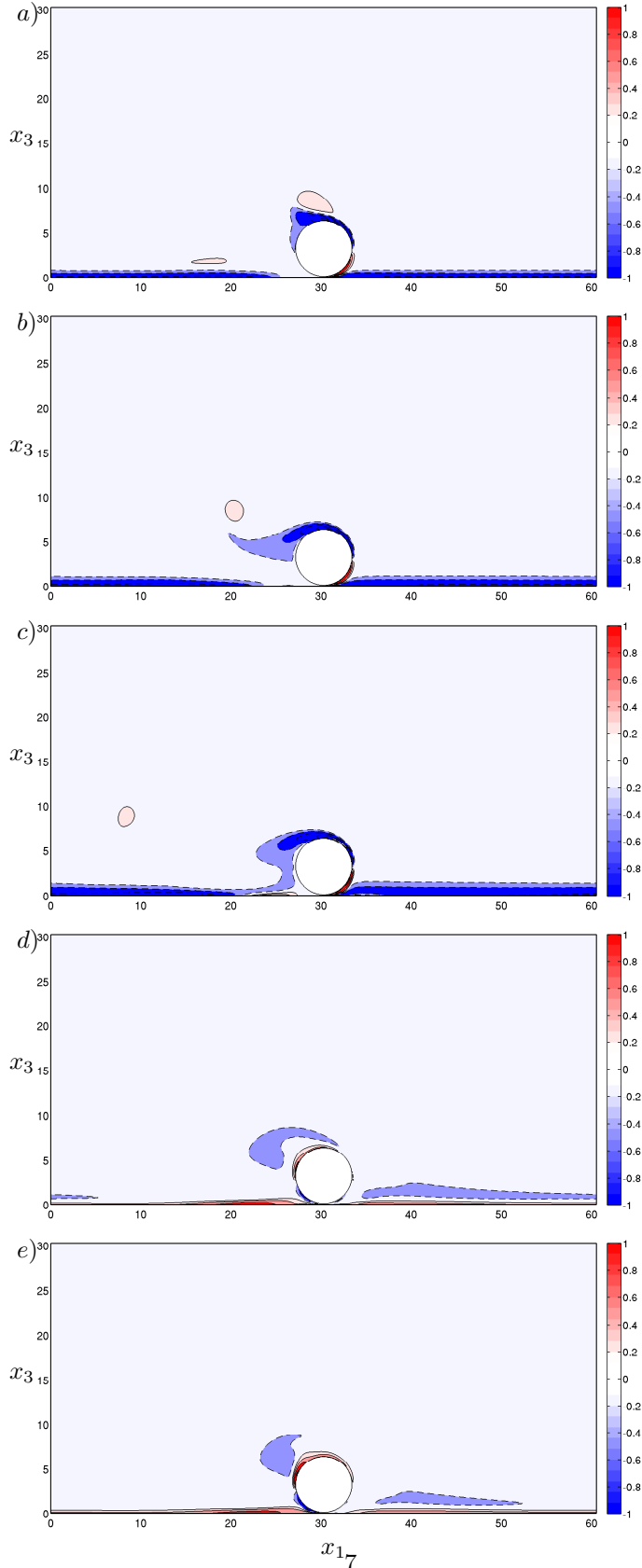
The bottom is assumed to be made up of spherical sediment grains of size $d^*$ resting on a plane wall located at 
$x^*_3 = 0$.
Because of numerical reasons, as in \citet{fischer2002}, the small spheres, which 
mimic the sediment grains, do not touch but they are $\epsilon^*$ apart 
from the bottom.
Then, a much larger spherical object, characterized by a diameter $D^*$, is laid down on the 
sediment bed (see figure \ref{fig1}).

The hydrodynamic problem, the solution of which describes the flow close to the bottom, is written in dimensionless form by introducing 
the following variables
\begin{equation}
\label{var}
t = t^* \omega^*; \ \ \ \ 
(x_1,x_2,x_3) = \frac{(x^*_1, x^*_2, x^*_3)}{\delta^*}; \ \ \ \ 
(u_1,u_2,u_3) = \frac{( u^*_1, u^*_2,u^*_3)}{U^*_0}; \ \ \ \ p = \frac{p^*}{\rho^*(U^*_0)^2}
\end{equation}
In (\ref{var}), $t^* $ is time, $u^*_1, u^*_2, u^*_3$ are the fluid velocity
components along the $x^*_1$-, $x^*_2$- and $x^*_3$-axes, respectively, and
$\delta^*=\sqrt{2\nu^*/\omega^*}$ is the conventional thickness of the viscous boundary layer
close to the bottom, $\nu^*$ being the kinematic viscosity of the fluid. 
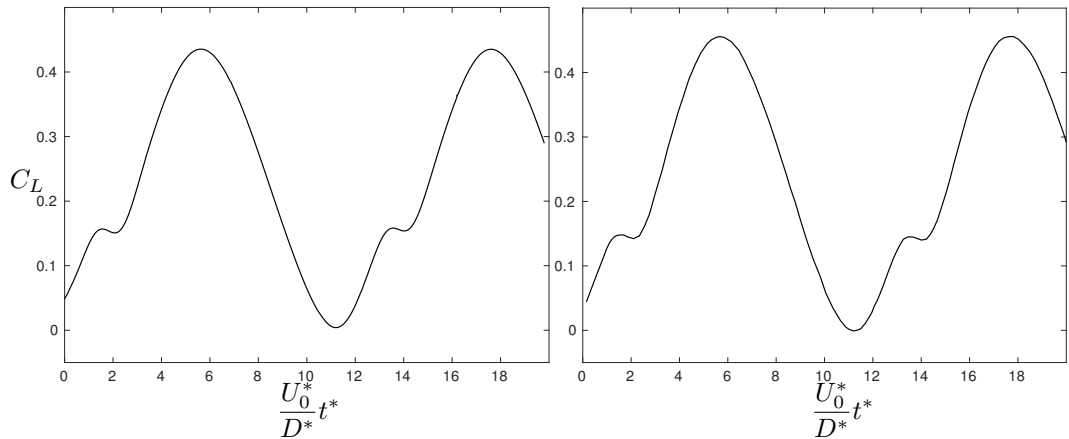
\begin{figure}
\begin{picture}(0,160)(0,0)
\ifthree
\put(-85,0){
\putpic{90,14}{.5}{figure3a}
\put(180,0){$\displaystyle\frac{U^*_0}{D^*} t^*$}
\put(80,87){$C_{L}$}
}
\put(55,0){
\putpic{144,14}{.5}{figure3b}
\put(240,0){$\displaystyle\frac{U^*_0}{D^*} t^*$}
}
\fi
\end{picture}
\caption{
Time development of the lift coefficient $C_L$ for a single sphere in an oscillatory boundary layer
over a smooth wall for $R_\delta=47.87$, $D=6.267$ and $\epsilon=\epsilon^*/\delta^*=0.09777$.
Left panel, present results; right panel, \citet{fischer2002}'s results.
}
\label{figB}
\end{figure}
Using the variables defined by (\ref{var}), continuity and Navier-Stokes equations read
\begin{equation}
\label{equ1}
\frac{\partial u_j }{\partial x_j}=0
\end{equation}
\begin{equation}
\label{equ2}
\frac{\partial u_i}{\partial t} + \frac{R_{\delta}}{2} \left(u_j \frac{\partial u_i}{\partial x_j} \right) = - \frac{R_{ \delta}}{2} 
\frac{\partial p}{\partial x_i} - \delta_{k1} \sin(t) + \frac{1}{2} 
\left( \frac{\partial^2 u_i }{\partial x_k \partial x_k} \right) + f_i
\end{equation}
where the Einstein's summation convention is used. Moreover, the Reynolds number $R_{\delta}$, which appears into (\ref{equ2}), is defined by
\begin{equation}
R_{\delta}=\frac{U_0^*\delta^*}{\nu^*}
\end{equation}
and the meaning of the terms $f_i$ is defined later on.

The continuity and momentum equations are solved numerically by means of a finite difference approach 
in a computational domain of dimensions $L_{x1}, L_{x2}$ and $L_{x3}$ in the streamwise, spanwise and vertical
directions, respectively. 
Equations (\ref{equ1})-(\ref{equ2}) are solved
throughout the whole computational domain, including the space occupied
by the sediment grains and the large spherical object, and appropriate force terms $f_i$ are added to the right hand side of (\ref{equ2})
to force the no-slip condition at the fluid-particle interfaces (immersed boundary approach).
In particular, the direct-forcing Immersed Boundary Method (IBM) proposed by \citet{uhlmann2005} is used to quantify the terms $f_i$ which are explicitly computed at each time step
as a function of the values of the velocity interpolated at nodes uniformly distributed on the spheres by means of the regularized delta function formulated by \citet{Roma1999}, without any feed-back procedure.

\vspace{.5cm}

Periodic boundary conditions are forced in the homogeneous directions ($x_1, x_2$), because the computational box is chosen large enough to include the largest 
vortex structures of the flow. At the upper boundary, located at $x_3=L_{x_3}$, the free stream condition 
is forced
\begin{equation}
\label{bc1x}
\left( \frac{\partial u_1}{\partial x_3},\frac{\partial u_2}{\partial x_3} \right)= (0,0) ; \ \ \ \  \ \ \ u_3=0 \ \ \ \ \ \  \mbox{at} \ \ \ x_3 = L_{x_3}
\end{equation}
which is equivalent to force the vanishing of the shear stresses, since at this elevation the flow
is assumed to be irrotational.
At the lower boundary of the fluid domain ($x_3=0$), where a rigid wall is located, the no-slip condition is forced
\begin{equation}
\label{bc1}
(u_1,u_2,u_3)= (0,0,0) \ \ \ \mbox{at} \ \ \ x_3 = 0
\end{equation}

The numerical approach solves the problem in primitive variables and uses
a fractional-step method to advance momentum equations in time.
A non-solenoidal intermediate velocity field is evaluated by means of momentum equations (\ref{equ1}) using a semi-implicit 
scheme of second order to discretize the viscous terms and a three-step, low-storage, 
self-restarting Runge-Kutta method to discretize explicitly the nonlinear terms. 
The implicit treatment of the viscous terms would require for the inversion of large sparse 
matrices which are reduced to three tridiagonal matrices by a factorization procedure with an error of order
$(\Delta t)^3$, $\Delta t$ being the time step of the numerical approach \citep{beam1976}. 
Then, by using momentum equation (\ref{equ2}) and forcing continuity equation (\ref{equ1}), a Poisson equation for the pressure field
is obtained, which is solved by means of an iterative procedure. Once the pressure field is obtained, the non-solenoidal velocity field is 
corrected to obtain a divergence-free velocity field.

An Adaptive Mesh Refinement (AMR) was used, which allows an oct-tree-structure local refinement 
of the grid in the regions of the flow, where large gradients are expected to be present. 
The use of the present adaptive mesh requires also the use of specific multi-grid solvers of
 the Helmholtz and Poisson problems which arise from the prediction step of the fractional-step scheme and
from the procedure used to evaluate of pseudo-pressure, respectively.
The Poisson solver implements the iterative procedure proposed by \citet{Ricker2008} which is based on 
the more general algorithm given by \citet{HuangGreengard2000}.
A similar approach was adopted to develop a multi-grid Helmholtz solver which exploits the 
alternate direction implicit approximation to generate, on the coarse mesh, an initial guess of the solution. 
Moreover, second order accurate interpolation/average operators were used at the interfaces between different refinement levels.
To guarantee the accuracy of the results, the grid spacing in the region close to the bottom
is chosen in the range $(0.056\delta^*,0.19\delta^*)$, depending on the parameters of the problem, and it increases up to a 
factor $8$ in the regions far from the bottom, where the flow is weakly influenced by the presence of the spheres. 
Table \ref{tab1} summarizes the values of the parameters of the numerical simulations, along with
some information on the size of the computational domain and the numerical grid. The reader should notice that the grid size,
which has the same value along the three spatial directions, is kept constant and equal to $\Delta x_{fine}$ in the regions characterized by the
presence of both the large sphere and the small spheres, in such a way that no adjustment of the IBM algorithm proposed by
\citet{uhlmann2005} is necessary.
A grid spacing, such that $d/\Delta x_{fine}\sim 10$, is used to guarantee a reliable description of the flow around the roughness elements.
Since the size of the computational domain is of order $\mathcal{O}(100\delta^*)$ and the time step is fixed in such a way that
the CFL number does not exceed $0.4$, remarkable computational resources are required by a typical run.
More details on the numerical procedure are described in \citet{Mazzuoli2016}.\\

\section{The results}

\subsection{Validation of the code}

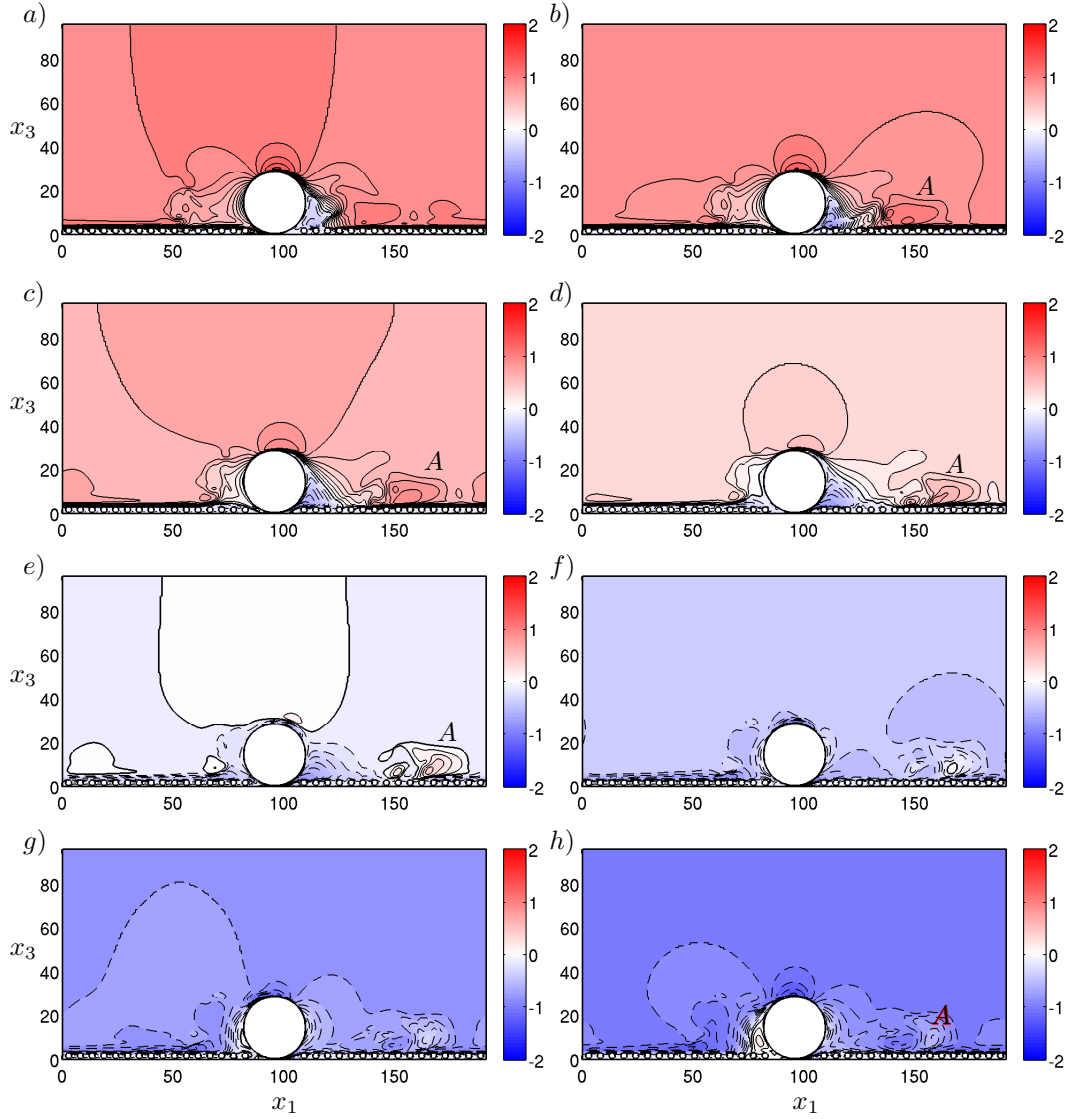
\begin{figure}
\begin{picture}(0,425)(0,0)
\iffour
\putpic{5,313}{.5}{figure4a}
\put(-5,368){$x_3$}
\putpic{200,313}{.5}{figure4b}
\putpic{5,208}{.5}{figure4c}
\put(-5,264){$x_3$}
\putpic{200,208}{.5}{figure4d}
\putpic{5,105}{.5}{figure4e}
\put(-5,161){$x_3$}
\putpic{200,105}{.5}{figure4f}
\putpic{5,2}{.5}{figure4g}
\put(-5,58){$x_3$}
\putpic{200,2}{.5}{figure4h}
\put(93,0){$x_1$}
\put(288,0){$x_1$}
\put(334,345){$A$}
\put(345,240){$A$}
\put(150,241){$A$}
\put(155,139){$A$}
\put(340,32){\cbr{$A$}}
\put(0,410){$a)$}
\put(197,410){$b)$}
\put(0,305){$c)$}
\put(197,305){$d)$}
\put(0,202){$e)$}
\put(197,202){$f)$}
\put(0,98){$g)$}
\put(197,98){$h)$}
\fi
\end{picture}
\caption{Streamwise velocity component at different phases of the cycle in the middle
vertical plane crossing the largest sphere 
for $R_\delta=56$, $D=28$, $d=2.6$ and $\epsilon=0.374$.
The thin solid ($u_1>0$) and broken ($u_1<0$) lines are the iso-velocity contours $( \Delta u_1=0.1 )$ and the thick
solid line corresponds to $u_1=0$. 
a) $t=\pi$, b) $t=9 \pi/8$, c) $t=10 \pi/8$, d) $t=11 \pi/8$, e) $t=12 \pi/8$, f) $t=13 \pi/8$, g) $t=14 \pi/8$, h) $t=15 \pi/8$.}
\label{fig2}
\end{figure}
\begin{figure}
\begin{picture}(0,440)(0,0)
\iffive
\putpic{2,282}{.5}{figure5a}
\putpic{200,282}{.5}{figure5b}
\putpic{2,135}{.5}{figure5c}
\putpic{200,135}{.5}{figure5d}
\putpic{2,3}{.5}{figure5e}
\putpic{200,3}{.5}{figure5f}
\put(283,0){$x_1$}
\put(83,0){$x_1$}
\put(-7,75){$x_3$}
\put(-7,215){$x_3$}
\put(-7,360){$x_3$}
\put(-2,425){$a)$}
\put(-2,275){$c)$}
\put(-2,128){$e)$}
\put(198,425){$b)$}
\put(198,275){$d)$}
\put(198,128){$f)$}
\fi
\end{picture}
\caption{Streamwise velocity component at $t=\pi$, when the free stream velocity is maximum, in the region close to the bottom 
for $D=28$, $d=2.6$, $\epsilon=0.374$ and $R_\delta=56$ (left hand side panels) and $R_\delta=112$ (right hand side panels).
a,b) Region on the downstream side; c,d) region on the upstream side and e,f) region far from the large sphere.
The thin solid ($u_1>0$) and broken ($u_1<0$) lines are the iso-velocity contours $( \Delta u_1=0.1 )$ 
and the thick solid line corresponds to $u_1=0$. 
}
\label{fig2a}
\end{figure}
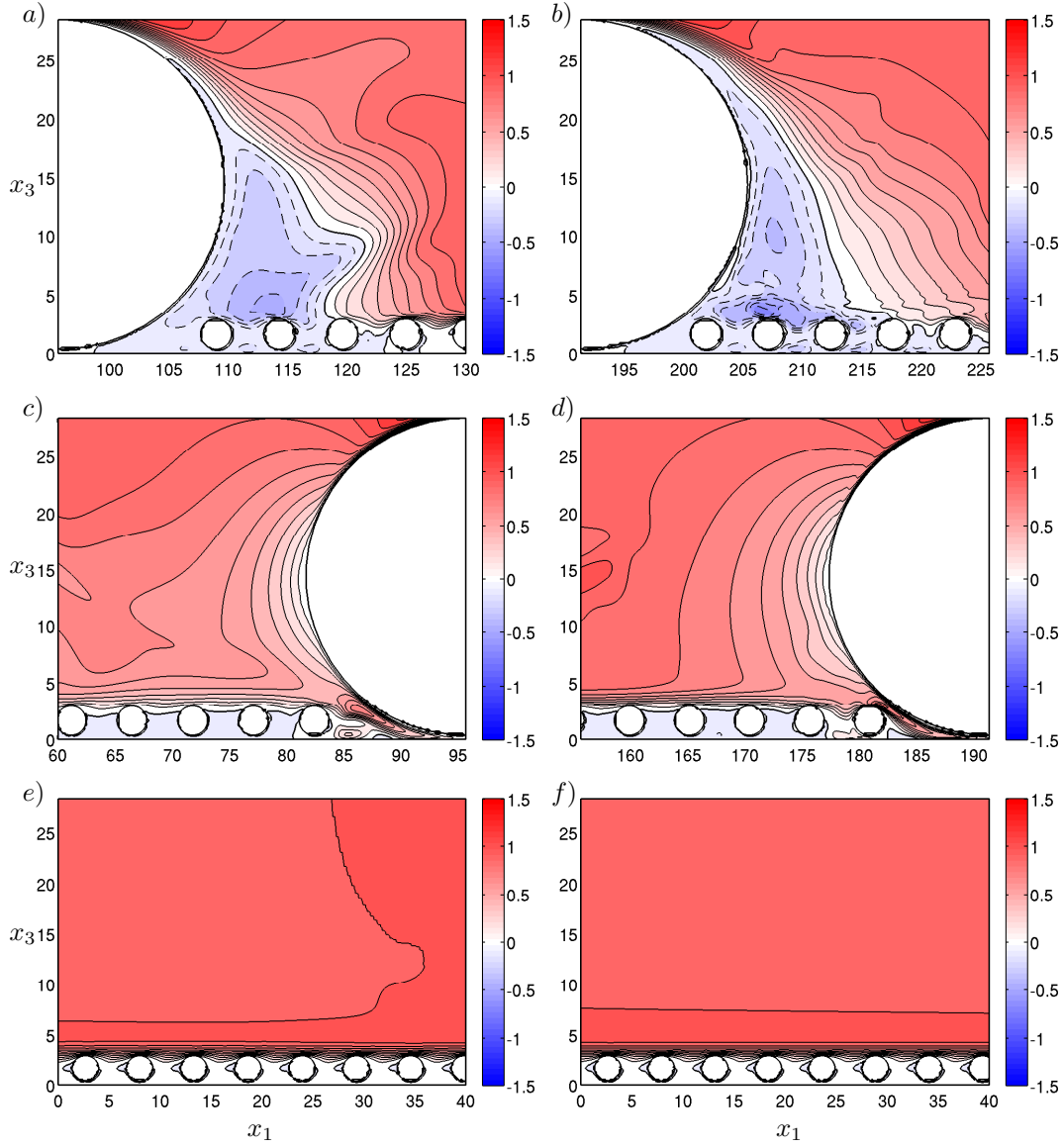
\begin{figure}
\begin{picture}(0,422)(0,0)
\ifsix
\putpic{5,314}{.5}{figure6a}
\put(-5,371){$x_2$}
\putpic{200,314}{.5}{figure6b}
\putpic{5,211}{.5}{figure6c}
\put(-5,268){$x_2$}
\putpic{200,211}{.5}{figure6d}
\putpic{5,108}{.5}{figure6e}
\put(-5,164){$x_2$}
\putpic{200,108}{.5}{figure6f}
\putpic{5,2}{.5}{figure6g}
\put(-5,58){$x_2$}
\putpic{200,2}{.5}{figure6h}
\put(92,0){$x_1$}
\put(287,0){$x_1$}
\put(0,410){$a)$}
\put(197,410){$b)$}
\put(0,307){$c)$}
\put(197,307){$d)$}
\put(0,204){$e)$}
\put(197,204){$f)$}
\put(0,98){$g)$}
\put(197,98){$h)$}
\put(122,354){\cbg{$C$}}
\put(122,382){\cbg{$B$}}
\put(327,354){\cbg{$C$}}
\put(327,382){\cbg{$B$}}
\put(157,144){
\put(85.5,29.5){
\put(40,0){\color{red}\line(1,0){26}}
\put(40,18){\color{red}\line(1,0){26}}
\put(40,0){\color{red}\line(0,1){18}}
\put(66,0){\color{red}\line(0,1){18}}
}}
\fi
\end{picture}
\caption{
Streamwise velocity component at different phases of the cycle in the horizontal plane crossing the largest sphere
for $R_\delta=56$, $D=28$, $d=2.6$ and $\epsilon=0.374$.
The thin solid ($u_1>0$) and broken ($u_1<0$) lines are the iso-velocity contours $( \Delta u_1=0.1 )$ and the thick
solid line corresponds to $u_1=0$. 
a) $t=\pi$, b) $t=9 \pi/8$, c) $t=10 \pi/8$, d) $t=11 \pi/8$, e) $t=12 \pi/8$, f) $t=13 \pi/8$, g) $t=14 \pi/8$, h) $t=15 \pi/8$.
}
\label{fig3}
\end{figure}
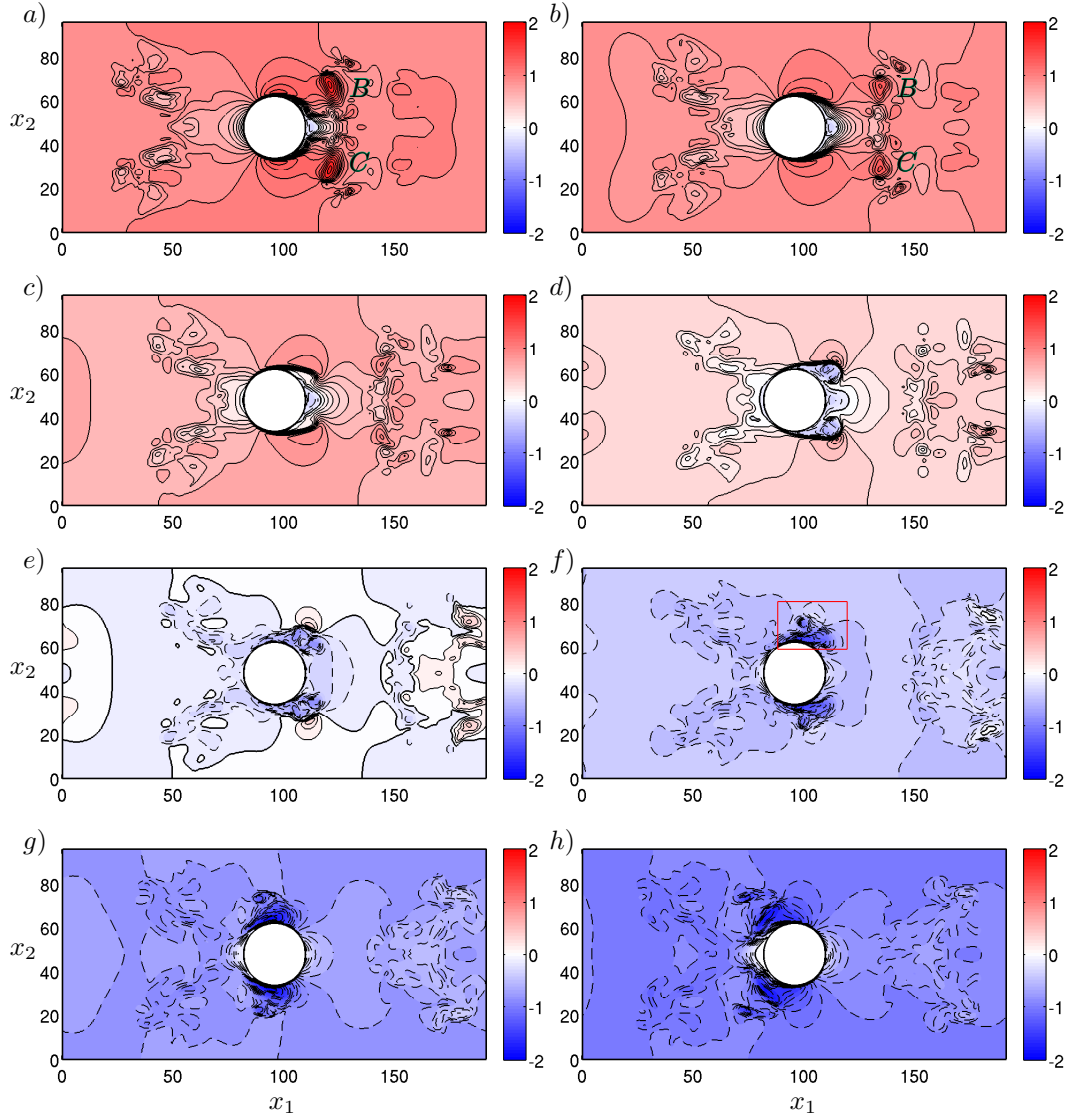
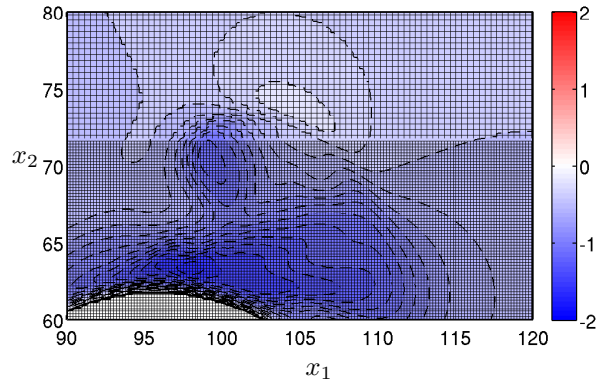
\begin{figure}
\begin{picture}(0,150)(0,0)
\ifsix
\put(90,0){
\putpic{0,0}{.55}{figure7}
\put(-10,78){$x_2$}
\put(100,-2){$x_1$}
}
\fi
\end{picture}
\caption{Detail of the panel $(f)$ of figure \ref{fig3} (red box). The computational grid is overlapped to contour patches (two confining mesh refinements can be seen). See the caption of figure \ref{fig3} for the values of the simulation parameters.}
\label{fig3aa}
\end{figure}

To validate the numerical approach, the oscillatory flow over a plane and smooth wall, where a single sphere is kept fixed,
was simulated for a value of the Reynolds number $R_\delta$ equal to $47.87$. The dimensionless diameter $D=D^*/\delta^*$ of the sphere was set equal to 
$6.267$ and the sphere did not touch the bottom but its dimensionless distance $\epsilon=\epsilon^*/\delta^*$ from it was set equal to $0.0978$,
which is equal to the value used by \citet{fischer2002}.
These values of the parameters were chosen to allow a comparison of the present results with those of \citet{fischer2002}.
Figure \ref{figA} shows the spanwise vorticity component in a vertical plane aligned
with the direction of the fluid oscillations and crossing the centre of the sphere. Different phases of the cycle are considered
after the flow reached a periodic state. The phases are chosen equal to those considered by \citet{fischer2002} in their figure 3
and a qualitative comparison can be easily made. 
This comparison shows that the present approach provides a reliable description of the oscillatory flow around an object.
Indeed, looking at figure \ref{figA} and at figure 3 of \citet{fischer2002}, it can be observed that
counter-clockwise spanwise vorticity is generated along the sphere surface when the external flow moves from the right to
the left of the figure. The production of counter-clockwise vorticity intensifies, when the clockwise vortex shed during the previous
half cycle is convected from the right to the left and travels above the top of the sphere. Then, the flow separates at the sphere 
crest and the free shear layer tends to be convected by the clockwise vortex which moves because dragged by the free stream flow.
Later, further counter-clockwise vorticity  is shed, which tends to roll-up and to generate a counter-clockwise rotating vortex.
Meanwhile, the clockwise rotating vortex, previously shed by the sphere surface, dissipates because of viscous effects.
Then, the counter-clockwise rotating vortex is convected from the left to the right and the phenomenon repeats specularly during the
following half cycle.

No quantitative comparison can be made between the vorticity computed by means of the present code and that described by \citet{fischer2002}, 
because \citet{fischer2002} do not provide the values of the isovorticity lines. A quantitative
estimate of the accuracy of the present results can be made by comparing the time development of the lift coefficient $C_{L}$
plotted in figure \ref{figB} with that drawn in figure 3 of \citet{fischer2002}. 
Herein, the lift coefficient is defined as the ratio between the lift force and the quantity $\frac{1}{8} \rho^* \pi D^{*2} U^{*2}_0$. 
The agreement turns out to be fair (the largest difference is a few percent of the maximum value) 
and similar to that observed comparing the velocity profiles with those obtained by \citet{fischer2002}.

\subsection{The velocity field}

Once the reliability of the code was ascertained, results were obtained by considering a large sphere over a rough bed
made by much smaller spheres. The first simulation is characterized by $R_\delta= 56.0, D=28.0$ and $d=2.6$. Moreover, because
of the numerical approach, the spheres do not touch the bottom but are $0.374\delta^*$ from it and are laid in a hexagonal arrangement. 
The distance of the spheres from the bottom, which is required for numerical reasons, is larger than that used in the previous run because small values of $\epsilon^*$
imply large computational costs and the value of
$\epsilon^*$ does not affect significantly the flow above the roughness elements. Indeed, the results of \citet{Fornarelli2009}
and those of \citet{Mazzuoli2016}, who computed the oscillatory flow above semispheres and spheres, respectively, show that
the geometry of the wall roughness below the sphere centres does not affect significantly the flow above them.
Figure \ref{fig2} shows the value of the streamwise velocity component plotted in the vertical mid-plane aligned with the direction
of the fluid oscillations, at different phases of the cycle
starting from $\pi$ to $15 \pi/8$ every $\pi/8$, once the flow field attained a periodic state.
As expected, the presence of the large sphere slows down the fluid in front of the sphere, while a recirculating region appears
behind it. Moreover, high velocities are present just above the crest of the sphere.
A careful analysis of the figure shows the presence of further high velocity regions in the wake of the sphere, the most evident of
which is denoted by the label $A$ in the figure. These regions, which are close to the bottom (see 
and compare figures \ref{fig2}b,c,d,e), move and suggest the existence of vortex structures which are convected by the oscillating 
fluid and feel the effects of the self-induced velocity due to the interaction of the vortices with their image vortices below the seafloor.
These vortex structures survive for a significant time interval as it can be inferred by the presence of regions (the most evident of which
is denoted by $A$ in figure \ref{fig2}h) characterized by a positive velocity, while the fluid far from the bottom has a significant negative velocity.

The smaller spheres act as a bottom roughness and create a layer of fluid, the thickness of which 
is $\mathcal{O}(d)$, where the velocity almost vanishes. To show the details of the flow around the small spheres, 
the left hand side panels of figure \ref{fig2a} show enlargements of the region close to the bottom, when the free stream velocity is positive and maximum.
Near the large sphere, on the downstream side (see figure \ref{fig2a}a), the streamwise 
velocity component close to the bottom is directed from the right to the left while the free stream velocity is in the opposite direction. 
The flow reverses its direction because of the clockwise vorticity shed by the large sphere which induces negative velocities close to the bottom. 
On the other hand, on the upstream side (see figure \ref{fig2a}c), the velocity is positive but it assumes small values because of the blockage 
effect due to the resting large sphere. However, around the base of the sphere, the fluid accelerates and the streamwise velocity increases 
because the pressure on the stoss side of the sphere is larger than the pressure on the lee side. At last, far from the sphere 
(see figure \ref{fig2a}e), the velocity field is 'homogeneous' in the streamwise and spanwise directions. In particular, within the gaps among 
the small spheres, the fluid is practically at rest and significant velocities are observed only above the crests of the spheres.
Figure \ref{fig2a} (right hand side panels) shows also results for the same values of the parameters but for $R_\delta=112$, which
are discussed later on.

Figure \ref{fig3} shows again the streamwise velocity component, at the same phases of the cycle as those considered in figure \ref{fig2}, 
but in a horizontal plane which crosses the centre of the large sphere. As expected, on the upstream side, the fluid slows down when 
approaching the sphere because of its blockage effect, while on the lateral sides it accelerates and the flow separates from the sphere surface 
along the downstream side, generating recirculating regions behind the sphere. Moreover, the presence of intense vortex structures shed by the sphere
can be inferred by the regions of high velocity denoted by $B$ and $C$ in figure \ref{fig3}a,b.

Some of the panels of figure \ref{fig3} show very small flow structures which might appear not fully resolved. To show that the numerical grid is small
enough to provide an accurate description of the time development of the flow field, figure \ref{fig3aa} shows an enlargement of the region highlighted in figure 
\ref{fig3}f, where the numerical grid is also plotted. Figure \ref{fig3aa} shows that the grid size is much smaller than the smallest flow structures and it is 
small enough to provide reliable description of the viscous boundary layer which develops on the sphere surface 
(more quantitative results which support this statement are described later when larger values of the Reynolds number are considered such that
smaller vortices are generated).

The flow described by figures \ref{fig2}, \ref{fig2a}, \ref{fig3} is practically the mirror image of that observed during the previous and following 
half cycles. This finding, along with the symmetry of the flow with respect to the vertical plane crossing the centre of the large sphere and aligned with
the flow direction, suggests that the flow is not turbulent. However, the signal of the velocity just behind the large sphere (see figure \ref{fig4}) 
shows that rapid fluctuations of the velocity are superimposed on much slower oscillations. The former are induced by the passage of the vortex 
structures shed by the large sphere while the latter are due to the oscillating pressure gradient which forces the fluid motion.

Since it is expected that an increase of the Reynolds number leads to stronger nonlinear effects and causes the generation  of a larger number of vortex structures
of smaller size, it might be that a chaotic flow appears and/or turbulence is triggered when the Reynolds number is increased. Indeed, \citet{blondeaux1991} 
and \citet{vittori1993}, who investigated the two-dimensional oscillatory flow above a rippled bed and around a circular cylinder, respectively, found that
the nonlinear interaction of coherent vortex structures gives rise to a chaotic flow
when the Reynolds number becomes larger than a critical value.

\begin{figure}
\begin{picture}(0,350)(0,0)
\ifseven
\putpic{0,180}{.49}{figure8a}
\putpic{210,180}{.49}{figure8b}
\putpic{100,5}{.49}{figure8c}
\put(95,172){$t$}
\put(305,172){$t$}
\put(195,-5){$t$}
\put(-8,265){$u_1$}
\put(200,265){$u_3$}
\put(90,85){$u_2$}
\put(-8,330){$a)$}
\put(200,330){$b)$}
\put(90,155){$c)$}
\fi
\end{picture}
\caption{
a) streamwise, b) wall-normal and c) spanwise components of the velocity at $(x_1, x_3)=(123.67,14.39)$ plotted versus time
for $R_\delta=56$, $D=28$, $d=2.6$, $\epsilon = 0.374$ and $x_2=54.83$ $[$---$\bigcirc$---$]$, $x_2=61.83$ $[$---$\square$---$]$, 
$x_2=68.83$ $[$---$\vartriangle$---$]$.}
\label{fig4}
\end{figure}
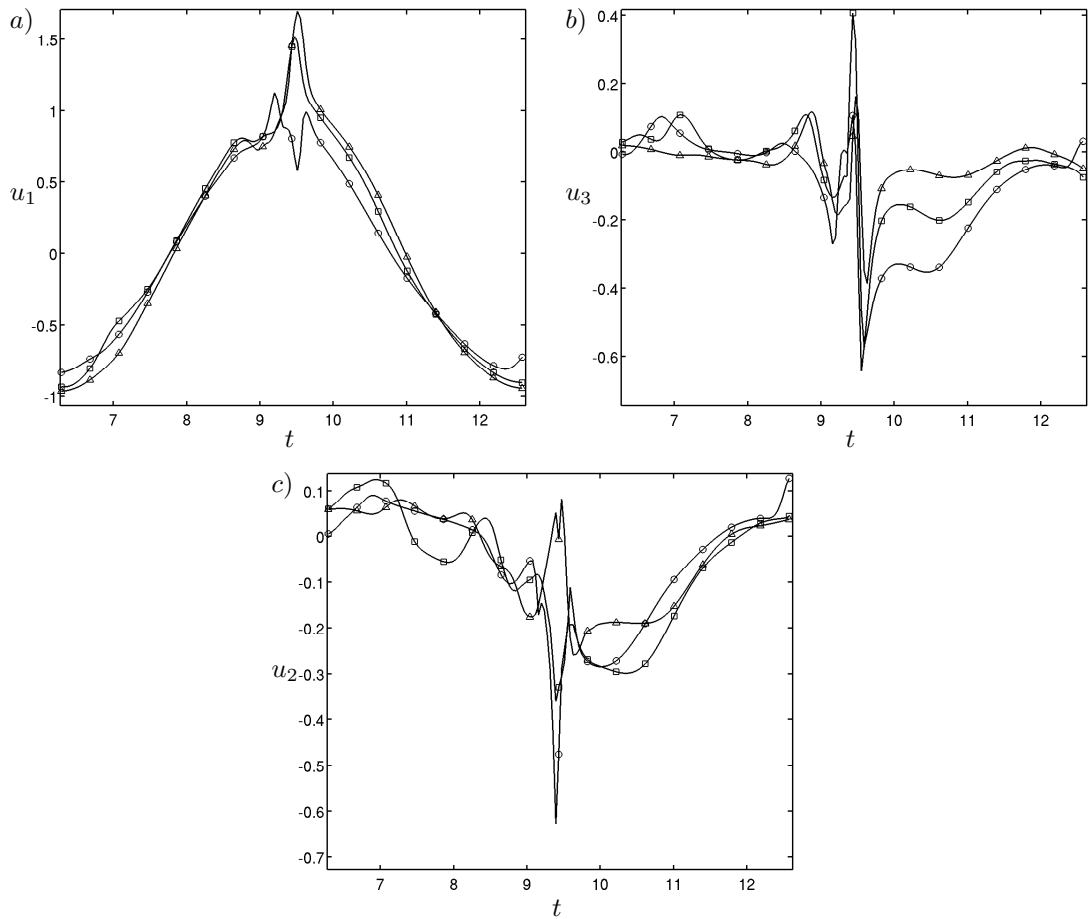
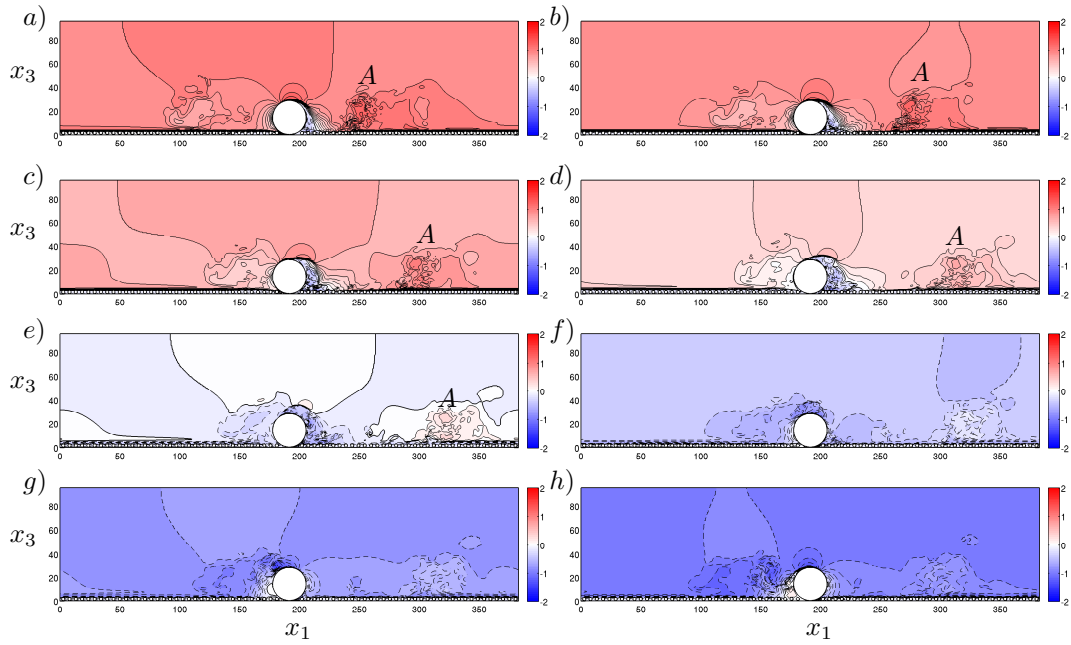
\begin{figure}
\begin{picture}(0,245)(0,0)
\iften
\putpic{5,178}{.5}{figure9a}
\put(-5,208){$x_3$}
\putpic{200,178}{.5}{figure9b}
\putpic{5,118}{.5}{figure9c}
\put(-5,149){$x_3$}
\putpic{200,118}{.5}{figure9d}
\putpic{5,60}{.5}{figure9e}
\put(-5,91){$x_3$}
\putpic{200,60}{.5}{figure9f}
\putpic{5,2}{.5}{figure9g}
\put(-5,33){$x_3$}
\putpic{200,2}{.5}{figure9h}
\put(98,-2){$x_1$}
\put(293,-2){$x_1$}
\put(125,206){$A$}
\put(332,207){$A$}
\put(345,145){$A$}
\put(147,146){$A$}
\put(155,84){$A$}
\put(0,228){$a)$}
\put(197,228){$b)$}
\put(0,168){$c)$}
\put(197,168){$d)$}
\put(0,110){$e)$}
\put(197,110){$f)$}
\put(0,53){$g)$}
\put(197,53){$h)$}
\fi
\end{picture}
\caption{Streamwise velocity component at different phases of the cycle in the middle
vertical plane crossing the largest sphere 
for $R_\delta=112$, $D=28$, $d=2.6$ and $\epsilon=0.374$.
The thin solid ($u_1>0$) and broken ($u_1<0$) lines are the iso-velocity contours $( \Delta u_1=0.1 )$ and the thick
solid line corresponds to $u_1=0$. 
a) $t=\pi$, b) $t=9 \pi/8$, c) $t=10 \pi/8$, d) $t=11 \pi/8$, e) $t=12 \pi/8$, f) $t=13 \pi/8$, g) $t=14 \pi/8$, h) $t=15 \pi/8$. 
}
\label{fig9}
\end{figure}
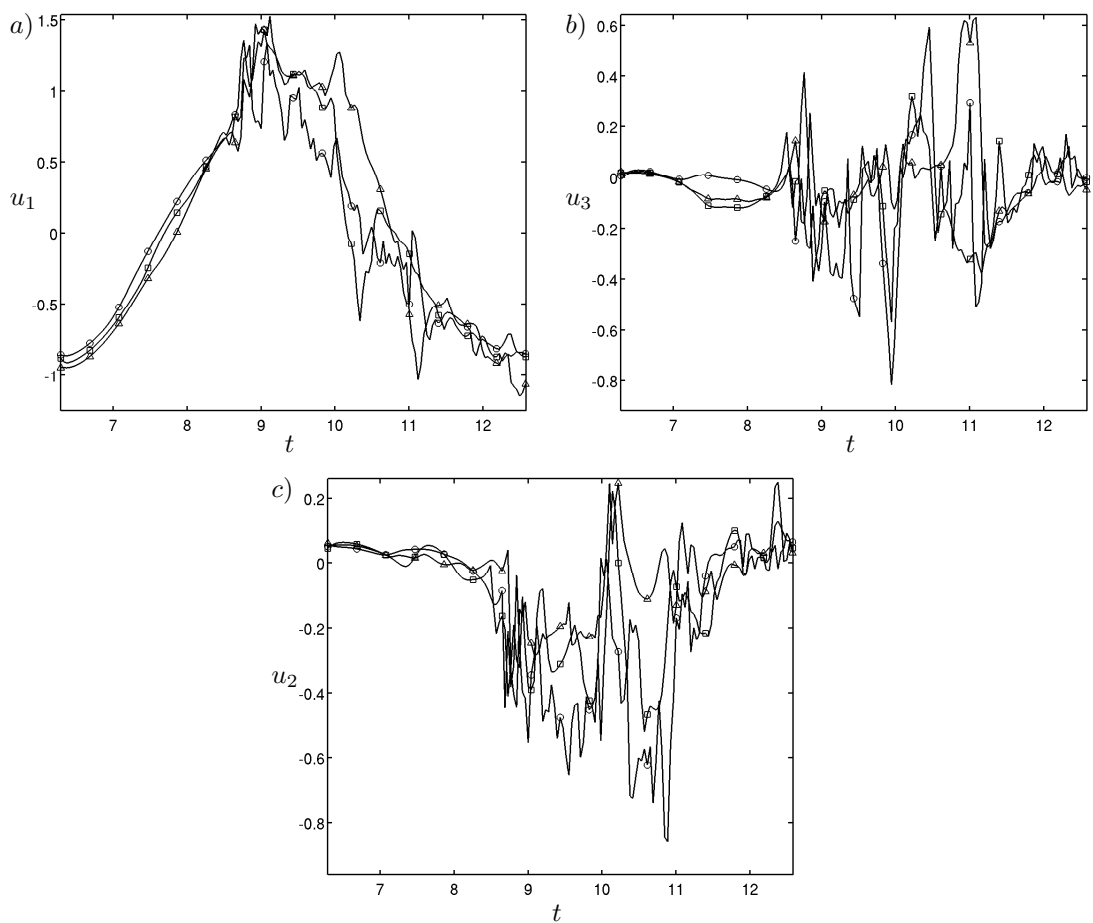
\begin{figure}
\begin{picture}(0,350)(0,0)
\ifeleven
\putpic{0,180}{.49}{figure10a}
\putpic{210,180}{.49}{figure10b}
\putpic{100,5}{.49}{figure10c}
\put(95,172){$t$}
\put(305,172){$t$}
\put(195,-5){$t$}
\put(-8,265){$u_1$}
\put(200,265){$u_3$}
\put(90,85){$u_2$}
\put(-8,330){$a)$}
\put(200,330){$b)$}
\put(90,155){$c)$}
\fi
\end{picture}
\caption{
a) streamwise, b) wall-normal and c) spanwise components of the velocity plotted versus time.
$x_1=123.67$, $x_3=14.39$ and $x_2=54.83$ $[$---$\bigcirc$---$]$, $x_2=61.83$ $[$---$\square$---$]$, $x_2=68.83$ $[$---$\vartriangle$---$]$. $R_\delta=112$, $D=28$, $d=2.6$, $\epsilon = 0.374$.}
\label{fig5}
\end{figure}

To ascertain whether an increase of the Reynolds number leads to a turbulent flow, a further run was made for the same values of the parameters
($D=28$, $d=2.6$, $\epsilon = 0.374$) but for $R_\delta=112$ and figure \ref{fig9}, which is similar to figure \ref{fig2}, shows the streamwise velocity component. 
Since the dimensionless diameters of large and small spheres are kept fixed, 
the increase of Reynolds number is due to a proportional increase of $U^*_0$. It follows that the Keulegan-Carpenter number of the 
phenomenon, $K_c=U_0^*/(\omega^*D^*)$, increases too and this explains why the streamwise size of computational box was increased.

The results show that the vortex structures, which are generated during each half-cycle by the roll-up of the shear layer shed by the surface 
of the large sphere, break and originate an ensemble of small vortices on the lee side of the sphere. 
The nonlinear self-interaction of the small vortices in the wake of the large sphere gives rise to a turbulent flow, as it appears from figure \ref{fig5} where the time development of the three velocity components just behind the sphere is plotted. 
Indeed, the rapid velocity fluctuations, which are superimposed on the slow oscillations induced by the pressure gradient, 
turn out to have a random character. Moreover, the velocity field is no longer the mirror image of the velocity fields which are observed
during the following or previous half-cycles and the instantaneous flow field loses its symmetry with respect to the vertical plane which crosses the centre
of the sphere and is aligned with the direction of the fluid oscillations.

The results of the numerical simulation for $R_\delta=112$ 
show that the flow in the wake of the large sphere is turbulent but 
the evaluation of turbulence characteristics is not simple. First of all, the
flow induced by the forcing pressure gradient is oscillating and turbulence
characteristics are time-dependent and confined within relatively small regions. 
Moreover, the turbulent fluctuations are neither homogeneous nor isotropic. 

In principle, the average flow field could be evaluated by using a phase 
average procedure and by taking advantage of some symmetries of the problem. 
However, the large computational costs did not allow us to simulate the 
flow for a large number of cycles. Since turbulence is observed in the wake 
of the sphere within relatively small regions, where turbulence characteristics
weakly depend on the horizontal coordinates but they strongly vary in the 
vertical direction, we evaluate the velocity fluctuations $(u'_1, u'_2, u'_3)$
by subtracting a spatial averaged flow field $(\overline u_1, \overline u_2, \overline u_3)$ 
from the actual values of the velocity field. 
The spatial average is performed over a volume which extends in the horizontal directions, covering the region where turbulence
is detected, but it is quite thin, having a thickness equal to a few grid cells.
Moreover, the turbulent velocity fluctuations are evaluated only at particular phases of the oscillatory cycle and far from the sphere,
when and where turbulence can be assumed to be independent on the horizontal coordinates. In particular,
the regions close to the sphere, where the coherent vortices shed by the sphere make turbulence characteristics
to depend on $x_1$ and $x_2$ are not considered. 

\begin{figure}
\begin{picture}(0,190)(0,0)
\ifnineteen
\putpic{60,2}{.6}{figure11}
\put(50,88){$R_{11}$}
\put(180,-4){$\boldsymbol{r}$}
\fi
\end{picture}
\caption{Normalized correlation function $R_{11}$ of the fluctuations of the streamwise velocity component, 
as a function of the streamwise and spanwise separations $r_1$ and $r_2$, respectively, 
obtained at $t=5.13$ considering a rectangular volume centered in $(x_1,x_2,x_3)=(276.33,47.64,16.63)$ 
of dimensions $(46.71\times 46.71\times 1.87)$ for $R_\delta=112$, $D=28$, $d=2.6$.
}
\label{fig17x}
\end{figure}
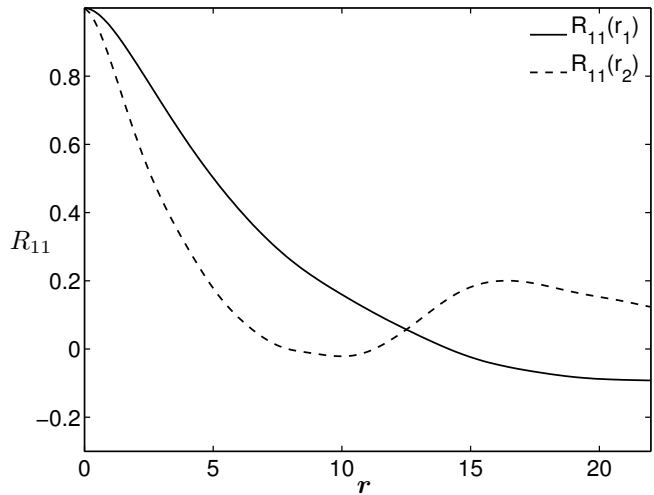

Then, the normalized two-point correlations of the velocity fluctuations are evaluated in the
streamwise and spanwise directions. The results allow one to gain an estimate of the size
of the vortex structures which characterize the turbulence field. 
For example, the normalized two-point correlation $R_{11}=\overline{u'_1({\bf x})u'_1({\bf x + r})}/\overline{u'_1({\bf x})u'_1({\bf x})}$
is plotted at $(x_1,x_2,x_3)=(276.33,47.64,16.63)$ as a function of $r_1$ and $r_2$ in figure \ref{fig17x}. The results 
allow an estimate of (i) the longitudinal and lateral integral length scales of the turbulent fluctuations, which turn out to be of order 
$10 \delta^*$, and (ii) longitudinal and lateral Taylor microscales, which turn out to be of order $\delta^*$.
It is worth pointing out that the evaluation of $R_{11}$ was not made for larger values of $r$, even though its
value does not vanish for $r=22$, because turbulence is present only in a small spatial region such that the evaluation of $R_{11}$
for larger values of $r$ would be meaningless.
At last, the dissipation rate $\varepsilon^* = \overline{(\partial u^*_i/\partial x^*_j)(\partial u^*_i/\partial x^*_j)}$ is evaluated
again at $t=5.13 \pi$ and it shows that the Kolmogorov scale $(\nu^{*3}/\varepsilon^*)^{1/4}$ is of order $10^{-1} \delta^*$,
thus indicating that the grid size is sufficiently small to provide reliable results on turbulence dynamics.
Turbulence is generated also close to the rough bottom by the interaction of the large scale coherent vortices, convected
by the free stream, with the roughness elements (the small spherical particles). 
The results show that in this region the velocity fluctuations are highly anisotropic. 
 
Even though turbulence is present in the wake of the large sphere, the bottom boundary layer (see Figure \ref{fig2a}f) and
the flow close to the large sphere (see Figure \ref{fig2a}b) keep in the laminar regime. 
Indeed, no random oscillations of the velocity field are present above the small spheres as long as they do not interact with the vortices 
originated from the large sphere (see Figure \ref{fig2a}f). 
This finding is in agreement with the results of \citet{Mazzuoli2016} who observed that
larger values of the Reynolds number are necessary to trigger turbulence appearance in the oscillatory boundary layer over
spherical particles of similar size. For example, \citet{Mazzuoli2016} found that the critical value of the Reynolds number 
for $d=2.32$ (a value which is close to $d=2.6$) ranges above $500$. 

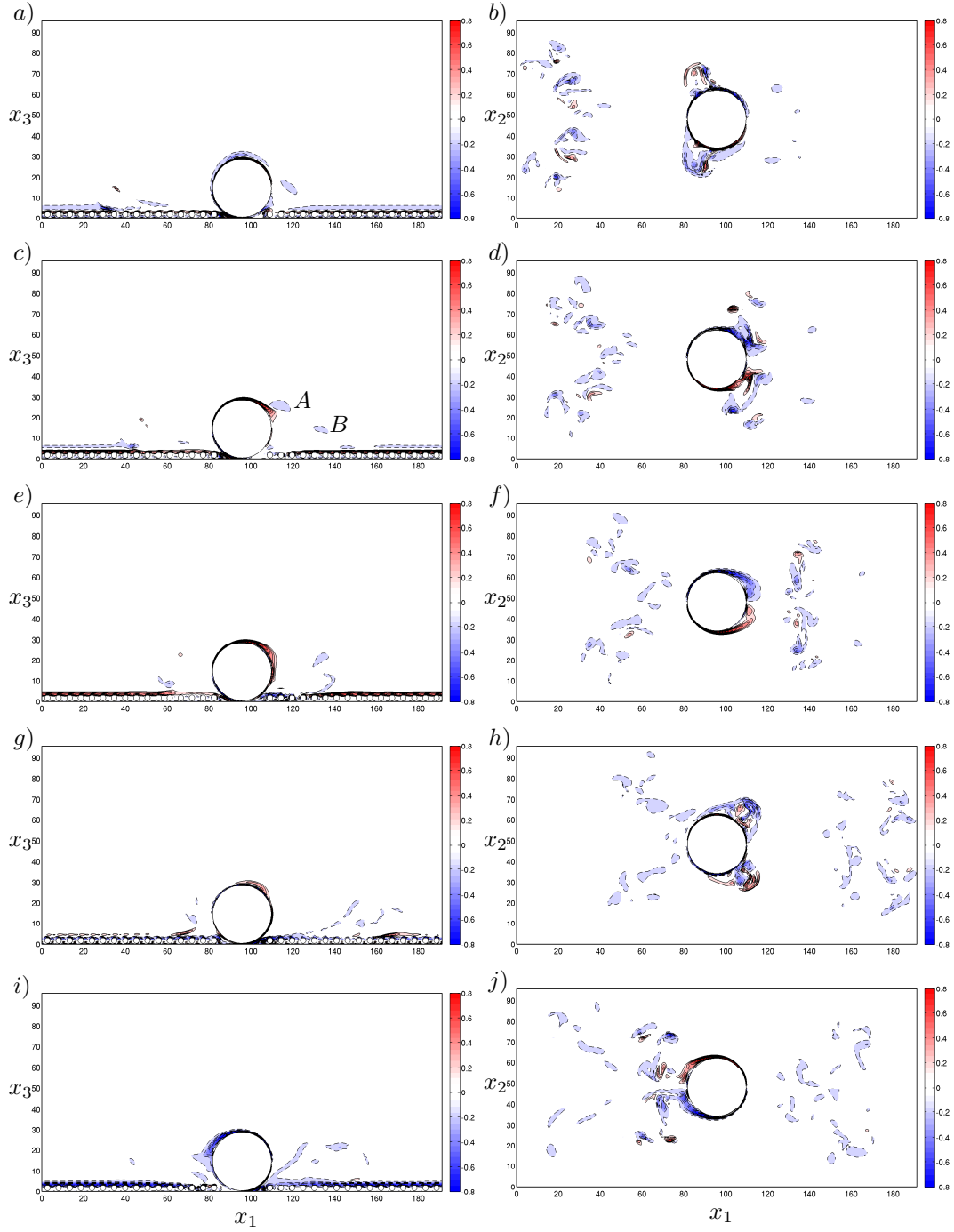
\begin{figure}
\begin{picture}(0,520)(0,0)
\iftwelve
\put(0,413){
\putpic{0,0}{.5}{figure12a}
\put(-5,50){$x_3$}
\put(-3,93){$a)$}
}
\put(0,311){
\putpic{0,0}{.5}{figure12c}
\put(-5,50){$x_3$}
\put(-3,93){$c)$}
\put(115,30){$A$}
\put(130,20){$B$}
}
\put(0,208){
\putpic{0,0}{.5}{figure12e}
\put(-5,50){$x_3$}
\put(-3,93){$e)$}
}
\put(0,105){
\putpic{0,0}{.5}{figure12g}
\put(-5,50){$x_3$}
\put(-3,93){$g)$}
}
\put(0,0){
\putpic{0,0}{.5}{figure12i}
\put(-5,50){$x_3$}
\put(90,-5){$x_1$}
\put(-3,93){$i)$}
}
\put(200,413){
\putpic{0,0}{.5}{figure12b}
\put(-5,50){$x_2$}
\put(-3,93){$b)$}
}
\put(200,311){
\putpic{0,0}{.5}{figure12d}
\put(-5,50){$x_2$}
\put(-3,93){$d)$}
}
\put(200,208){
\putpic{0,0}{.5}{figure12f}
\put(-5,50){$x_2$}
\put(-3,93){$f)$}
}
\put(200,105){
\putpic{0,0}{.5}{figure12h}
\put(-5,50){$x_2$}
\put(-3,93){$h)$}
}
\put(200,2){
\putpic{0,0}{.5}{figure12j}
\put(-5,50){$x_2$}
\put(90,-5){$x_1$}
\put(-3,93){$j)$}
}
\fi
\end{picture}
\caption{Spanwise component of the vorticity computed in the middle vertical plane 
(left hand side panels) and vertical component of the vorticity computed in the horizontal plane crossing the centre of the sphere (right hand side panels) 
for $R_\delta=56$, $D=28$, $d=2.6$ and $\epsilon=0.374$ at $t = 21/8\pi$, panels 
(a,b); $t=23/8\pi$, panels (c,d); $t=25/8\pi$, panels (e,f); $t=28/8\pi$, panels (g,h); $t=4\pi$, panels (i,j).
The vorticity isolines are equispaced by $\Delta \omega_2=0.1$ in the range $\pm 0.8$ but $\omega_2=0$; 
the value $\omega_2=-0.05$ is also considered (solid lines = positive values, broken lines = negative values).}
\label{fig10}
\end{figure}
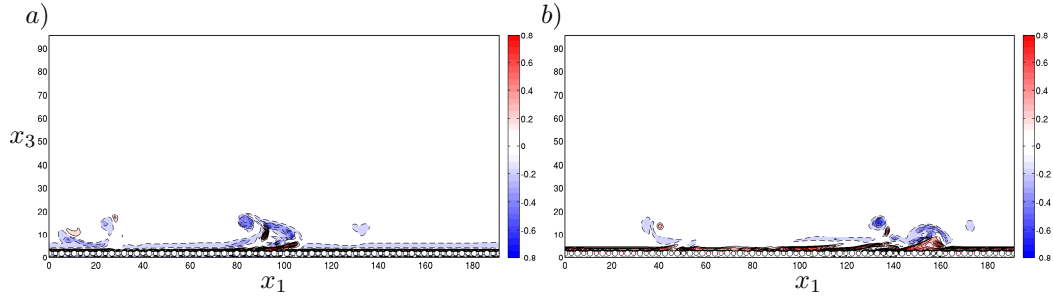
\begin{figure}
\begin{picture}(0,110)(0,0)
\iffifteen
\putpic{2,3}{.5}{figure13a}
\putpic{195,3}{.5}{figure13b}
\put(290,0){$x_1$}
\put(90,0){$x_1$}
\put(-4,55){$x_3$}
\put(2,100){$a)$}
\put(195,100){$b)$}
\fi
\end{picture}
\caption{Spanwise component of the vorticity computed in the plane $x_2 = x_{2c}+20$ ($x_{2c}$ being the spanwise 
coordinate of the centre of the large sphere) for $R_\delta=56$, $D=28$, $d=2.6$ and $\epsilon=0.374$ at 
$t = 21/8\pi$, panel (a) and $t=25/8\pi$, panel (b).
The vorticity isolines are equispaced by $\Delta \omega_2=0.1$ in the range $\pm 0.8$ but $\omega_2=0$; 
the value $\omega_2=-0.05$ is also considered (solid lines = positive values, broken lines = negative values).
}
\label{fig13}
\end{figure}
\begin{figure}
\begin{picture}(0,430)(0,0)
\iffourteen
\putpic{18,287}{.41}{figure14a}
\putpic{200,287}{.41}{figure14b}
\putpic{18,145}{.41}{figure14c}
\putpic{200,145}{.41}{figure14d}
\putpic{18,3}{.41}{figure14e}
\putpic{200,3}{.41}{figure14f}
\put(272,0){$x_2$}
\put(90,0){$x_2$}
\put(12,75){$x_3$}
\put(12,215){$x_3$}
\put(12,360){$x_3$}
\put(15,422){$a)$}
\put(195,422){$b)$}
\put(15,279){$c)$}
\put(195,279){$d)$}
\put(15,138){$e)$}
\put(195,138){$f)$}
\fi
\end{picture}
\caption{Streamwise component of the vorticity computed $a,b)$ in the middle vertical plane crossing the sphere 
center $x_1=x_{1c}=95.66$, $c,d)$ at $x_1=x_{1c}+0.5\:D$ and $e,f)$ at $x_1=x_{1c}+1.5\:D$ 
for $D=28$, $d=2.6$, $\epsilon=0.374$, $t = \omega^* t^*= 3\pi$. 
Left hand side panels, $R_\delta=56$ and right hand side panels, $R_\delta=112$.
The vorticity isolines are equispaced by $\Delta \omega_2=0.1$ in the range $\pm 0.8$ but $\omega_2=0$; the value $\omega_2=-0.05$ 
is also considered (solid lines = positive values, broken lines = negative values).
}
\label{fig12}
\end{figure}
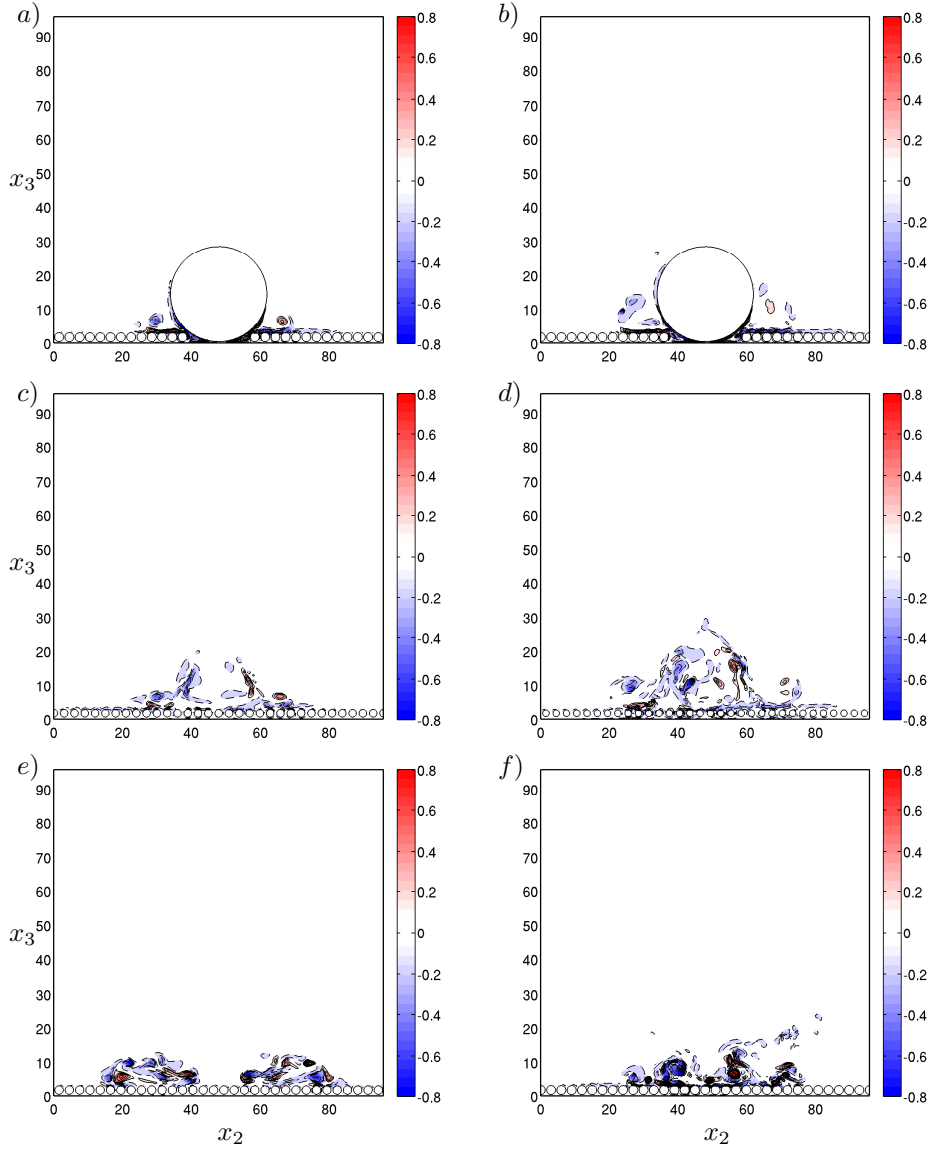

\subsection{The vorticity field and the dynamics of the vortex structures}

As discussed in the introduction, one of the aims of the work is the identification of the coherent vortex structures shed by the large
sphere and their influence on the forces exerted by the fluid on the sediment spherical particles resting on the bottom. 
Similarly to figure \ref{figA}, figure \ref{fig10} (left hand side panels) shows the spanwise vorticity component 
in the middle vertical plane crossing the sphere ($x_2 = 30$) and aligned with the direction of the fluid
oscillations for $R_\delta=56$, $D=28$, $d=2.6$ and $\epsilon=0.374$ for a few phases of the cycle. The main differences between the results of figure \ref{fig10} and those
of figure \ref{figA} are due to the different diameter of the large sphere and the presence of the spherical roughness elements in the former
case. The numerical simulation shows that clockwise vorticity is generated along the surface of the large sphere 
when the fluids moves from left to right (see figure \ref{fig10}a,c,e). 
Moreover, the production of vorticity increases as the fluid velocity close to the sphere increases because of either the external pressure 
gradient or the interaction of the sphere with a coherent vortex structure. 
When the flow reverses its direction, no further clockwise vorticity is generated
along the surface of the sphere (figure \ref{fig10}i) but the counter-clockwise rotating vortices (see vortices $A', B'$
in figure \ref{fig10}c) still moves from left to right because of their interaction with the seafloor (see figure \ref{fig10}e,g).
Figure \ref{fig10} shows that significant vorticity is
also shed close to the seafloor by the small spheres and a thick boundary layer is present close to the bottom.

Up to now, the vorticity dynamics is discussed looking at the flow in a vertical plane crossing the centre of the 
large sphere and aligned with the direction of the fluid oscillations. However, the flow is not two-dimensional. 
Indeed, the vorticity is more intense around the large sphere on the lateral sides where a strong interaction of the vortex
structures shed by the sphere takes place with the bottom roughness. In particular, figure \ref{fig13} shows that the local deceleration
of the external flow induced close to the bottom by a coherent vortex causes the separation of the bottom boundary layer and a complex
dynamics of the vorticity field.
The three-dimensional features of the flow are clearly shown in figure \ref{fig10} 
(right hand side panels) where the vertical vorticity component in a horizontal plane crossing the centre of the large sphere is plotted.
The vortex structures shed by the sphere during half cycle have a size larger than that which can be guessed on the basis of the spanwise vorticity plots
and they are characterized by a more complex
dynamics. In particular, figures \ref{fig10}b,d,h show that complex vortex structures exist where clockwise and counter-clockwise
vorticity strongly interact. These vortex structures move away from the sphere till they break and dissipate. 
Finally, figure \ref{fig12} shows the streamwise vorticity component 
in three different vertical planes orthogonal to the direction of the fluid oscillations and in the wake of the sphere, when the free stream velocity
is maximum ($t=3\pi$). The first plane crosses the centre of the large sphere, 
the second is tangent to the sphere on the downstream side and the third is one diameter apart. The figure shows that two counter-rotating 
streamwise vortices are present behind the sphere, which are close to the bed and strongly interact with the seafloor. 
In particular the streamwise vortices, which are close to the seafloor when they are near the sphere, lift from the bed as they
move away from the sphere.

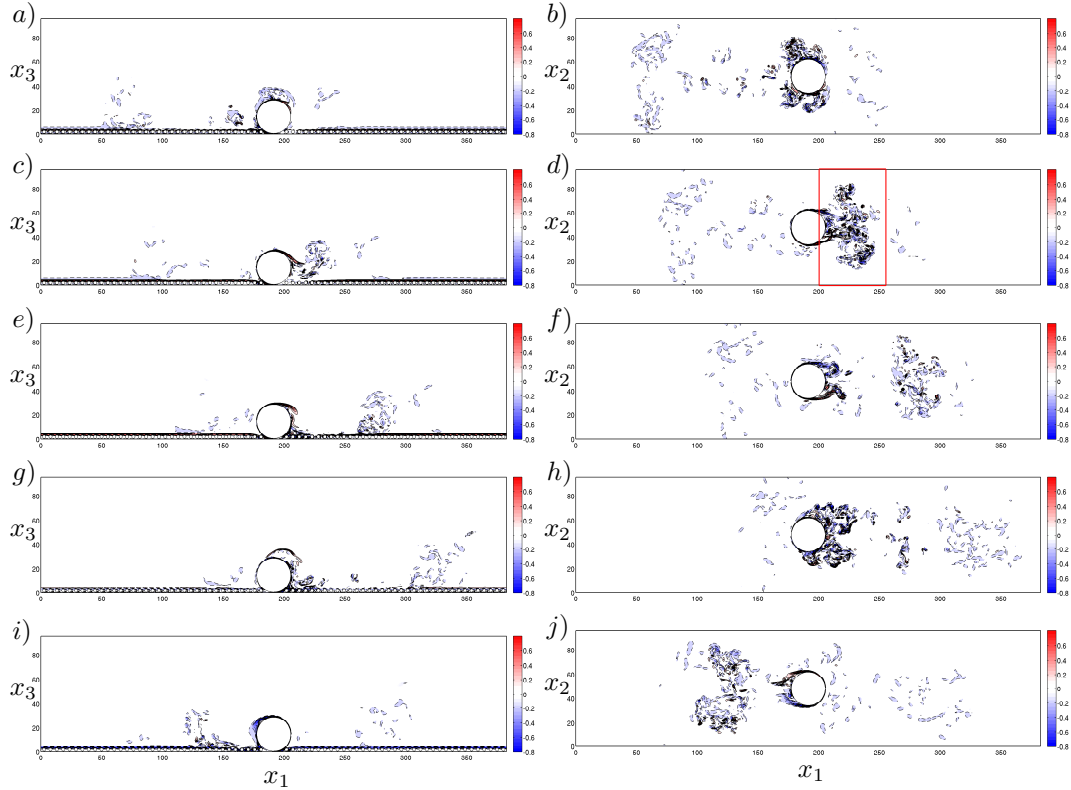
\begin{figure}
\begin{picture}(0,300)(0,0)
\iftwentyone
\put(0,233){
\putpic{0,0}{.5}{figure15a}
\put(-5,28){$x_3$}
\put(-4,49){$a)$}
}
\put(0,176){
\putpic{0,0}{.5}{figure15c}
\put(-5,28){$x_3$}
\put(-4,49){$c)$}
}
\put(0,118){
\putpic{0,0}{.5}{figure15e}
\put(-5,28){$x_3$}
\put(-4,49){$e)$}
}
\put(0,60){
\putpic{0,0}{.5}{figure15g}
\put(-5,28){$x_3$}
\put(-4,49){$g)$}
}
\put(0,0){
\putpic{0,0}{.5}{figure15i}
\put(-5,28){$x_3$}
\put(90,-5){$x_1$}
\put(-4,49){$i)$}
}
\put(200,233){
\putpic{0,0}{.5}{figure15b}
\put(-5,28){$x_2$}
\put(-4,49){$b)$}
}
\put(200,176){
\putpic{0,0}{.5}{figure15d}
\put(-5,28){$x_2$}
\put(-4,49){$d)$}
}
\put(200,118){
\putpic{0,0}{.5}{figure15f}
\put(-5,28){$x_2$}
\put(-4,49){$f)$}
\put(98,64.5){
\put(0,0){\color{red}\line(1,0){25}}
\put(0,44){\color{red}\line(1,0){25}}
\put(0,0){\color{red}\line(0,1){44}}
\put(25,0){\color{red}\line(0,1){44}}
}
}
\put(200,60){
\putpic{0,0}{.5}{figure15h}
\put(-5,28){$x_2$}
\put(-4,49){$h)$}
}
\put(200,2){
\putpic{0,0}{.5}{figure15j}
\put(-5,28){$x_2$}
\put(90,-5){$x_1$}
\put(-4,49){$j)$}
}
\fi
\end{picture}
\caption{
Spanwise component of the vorticity computed in the middle vertical 
plane (left hand side panels) and vertical component of the vorticity computed 
in the horizontal plane crossing the centre of the sphere (right hand side panels) 
for $R_\delta=112$, $D=28$, $d=2.6$ and $\epsilon=0.374$ at $t = 21/8\pi$, 
panels (a,b); $t=23/8\pi$, panels (c,d); $t=25/8\pi$, panels (e,f); $t=28/8\pi$, 
panels (g,h); $t=4\pi$, panels (i,j). 
The vorticity isolines are equispaced by $\Delta \omega_2=0.1$ and visualized in 
the range $\pm 0.8$ but $\omega_2=0$; the value $\omega_2=-0.05$ is also considered. 
(solid lines = positive values, broken lines = negative values).
}
\label{fig11}
\end{figure}
\begin{figure}
\begin{picture}(0,280)(0,0)
\ifthirteen
\put(-4,0){
\put(6,11){\includegraphics[trim=0cm 0cm 0cm 0cm,clip,width=.40\textwidth]{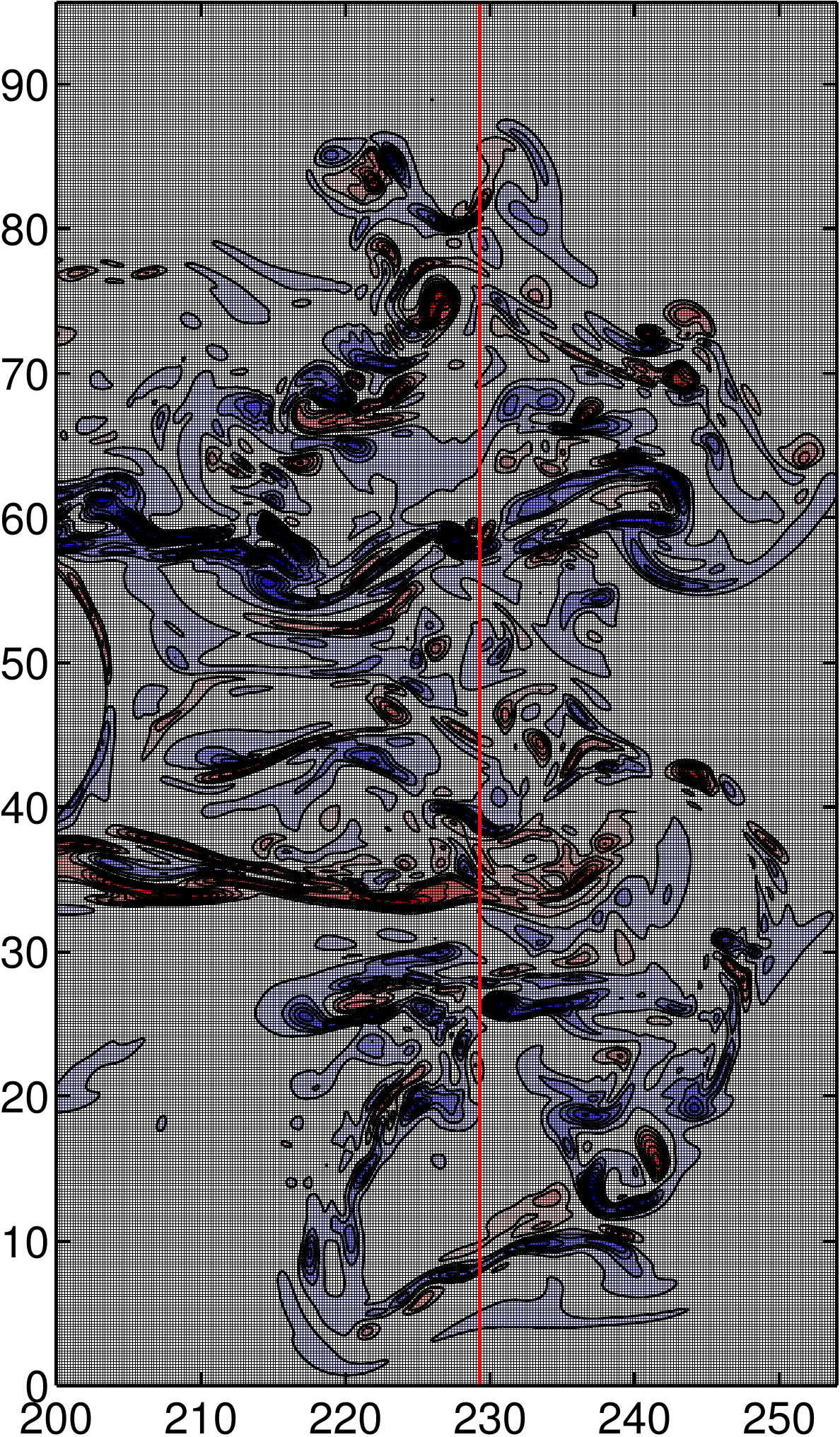}}
\put(-8,145){$x_2$}
\put(80,0){$x_1$}
\put(-5,260){$a)$}
\put(90,4){\color{red!}\small$A$}
\put(90,272){\color{red!}\small$B$}
}
\put(160,-8){
\putpic{0,3}{.6}{figure16b}
\put(40,160){$b)$}
}
\fi
\end{picture}
\caption{$(a)$ Spanwise component of the vorticity computed at $x_3=7.38$ inside the red box indicated the panel $(d)$ of figure \ref{fig11}. The computational grid is overlapped to the contour patches. 
$(b)$ Energy spectrum $E(\kappa_2)$ of the flow field shown in panel $(a)$ (red line A-B) is plotted versus the spanwise wavenumber $k_2$ for $x_1=229.3$.} 
\label{fig11a}
\end{figure}

A similar vorticity dynamics is observed for the larger value of $R_\delta$ ($R_\delta=112$), even though the vortex structures, shed by the
large sphere and moving away from it, break earlier and give rise to smaller vortices (see figure \ref{fig11}a-f).
The size of the turbulent eddies appearing into figure \ref{fig11} is quite small and the reader might question the reliability and accuracy of the results.
To show that also in this case the flow is well resolved, figure \ref{fig11a}a
shows an enlargement of the region highlightened in figure \ref{fig11}d along with the numerical grid which appears to be small enough to
provide reliable and accurate results. 
To quantitatively support this statement, the energy spectrum of the flow field 
shown in figure \ref{fig11a}a 
is computed as a function of the 
spanwise wavenumber $\kappa_2$ for $x_1=229.3$ and plotted in figure \ref{fig11a}b. The results show that the energy of the smallest resolved flow components
is negligible. Since similar results are obtained for different values of $(x_1, x_3)$ and $t$ (not shown herein), it clearly appears that the flow field is
well resolved.

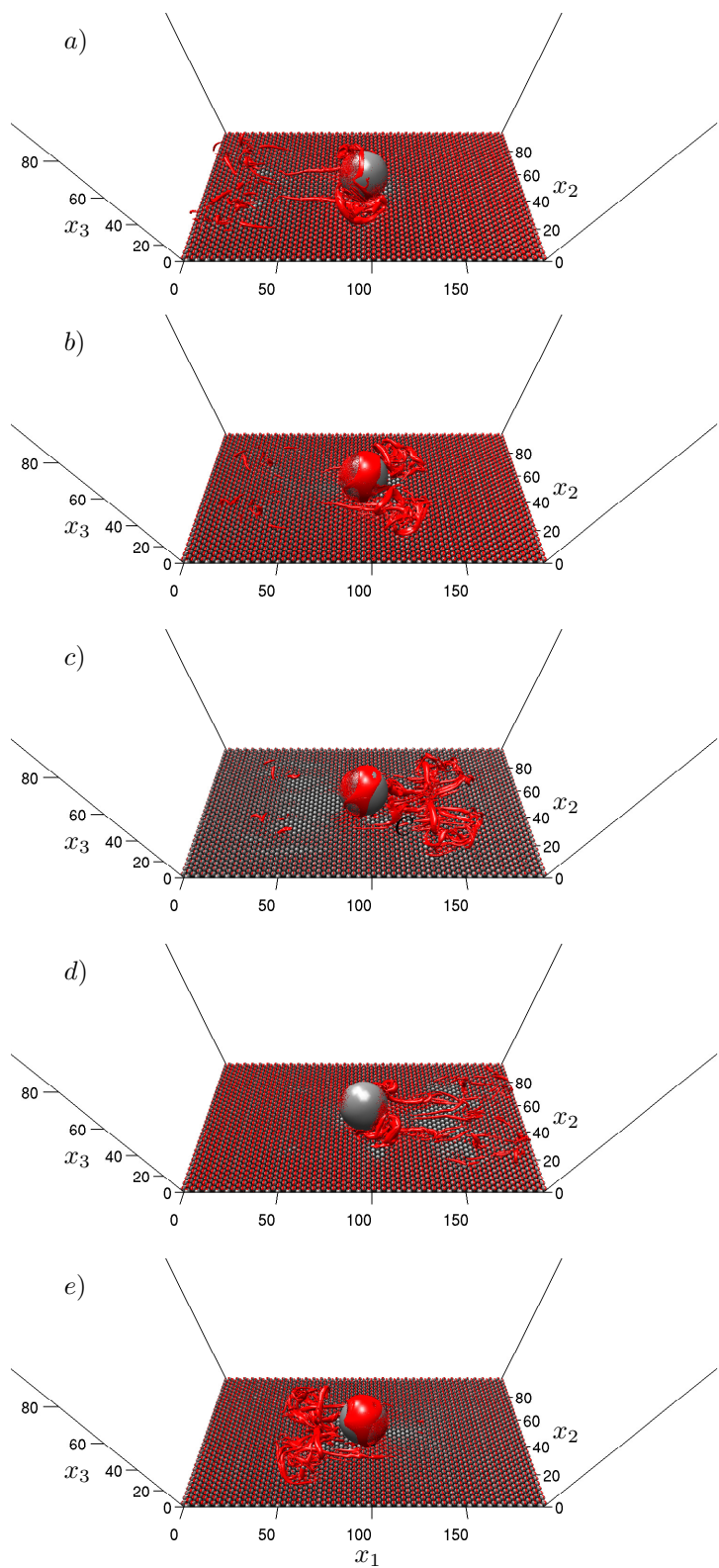
\begin{figure}
\begin{picture}(0,600)(0,0)
\ifeight
\putpic{60,480}{.7}{figure17a}
\putpic{60,365}{.7}{figure17b}
\putpic{60,245}{.7}{figure17c}
\putpic{60,125}{.7}{figure17d}
\putpic{60,5}{.7}{figure17e}
\put(190,-2){$x_1$}
\put(80,30){$x_3$}
\put(80,150){$x_3$}
\put(80,270){$x_3$}
\put(80,390){$x_3$}
\put(80,505){$x_3$}
\put(265,45){$x_2$}
\put(265,165){$x_2$}
\put(265,285){$x_2$}
\put(265,405){$x_2$}
\put(265,520){$x_2$}
\put(80,575){$a)$}
\put(80,460){$b)$}
\put(80,340){$c)$}
\put(80,220){$d)$}
\put(80,100){$e)$}
\put(205,275){$C$}
\fi
\end{picture}
\caption{Vortex structures visualized by (red) the contour surfaces of $\lambda_2=-0.1$, for $D=28, d=2.6, R_\delta=56$ at (a)~$t = 21\pi/8$, (b) $t=23\pi/8$, (c) $t=25\pi/8$, (d) $t=28\pi/8$, (e) $t=4\pi$.}
\label{fig6}
\end{figure}

Since the analysis of isosurfaces of the velocity and vorticity components provides a partial information on the coherent vortex structures 
and may be considered inadequate to detect vortices in an unsteady three-dimensional flow, we computed the eigenvalues of the symmetric tensor 
$\overline D^2+\overline \Omega^2$ ($\overline D$ is the strain rate tensor and $\overline \Omega$ is the spin tensor) and we considered the 
regions with two negative eigenvalues. Indeed, as discussed by \citet{JeongHussain1995}, these regions correlate well with coherent
vortex structures buried in a background vorticity field. Figure \ref{fig6} visualizes the isosurface characterized by a negative value 
of $\lambda_2$, which is the second eigenvalue of the tensor $\overline D^2+\overline \Omega^2$ ($\lambda_2=-0.1$) at different phases of the cycle for $R_\delta=56$.
The results plotted in figure \ref{fig6} allow to visualise the three-dimensional structure of the vorticity field and to gain a clear idea
of its dynamics. In particular, at $t=21\pi/8$ (figure \ref{fig6}a), the fluid moves from the right to the left and the roll-up of the vorticity shed by the sphere 
generates two vortex structures at the base of the sphere which grow in time because of the countinuous shedding of vorticity.
Later, when the flow reverses its direction, these vortex structures no longer grow but are simply convected from the left to the right.
In particular, when the vortex structures come close to the sphere, they strongly interact with it and induce the shedding of two new vortex structures which couple
with the vortices previously released from the sphere and move away from the sphere because of the free stream flow and the self-induced velocity (see figure \ref{fig6}b).
Meanwhile, two further vortex structures are shed by the sphere while the vortices previously generated decay because of viscous effects and their
interaction with the rough bottom (see figure \ref{fig6}d). The vorticity dynamics during the following half cycle is practically the mirror image of that previously described.
It is interesting to point out that those, which appear as vortex structures generated by the roll-up of vorticity of the same sign,
are indeed generated by the strong interaction of vortex structures of different sign as it can be inferred by the results plotted in 
figure \ref{fig10}d, where it appears clearly that the curved vortex structures which move away from the sphere (denoted as $C$ in figure 
\ref{fig6}c are due to both clockwise and counter-clockwise vorticity (see figure \ref{fig10}d,f).

\subsection{The bottom shear stress and the forces acting on bottom particles}

The present investigation was also aimed at identifying the regions of the seafloor where the vortices shed by the larger sphere tend to
set the sediments into motion and sweep them away. A simple analysis shows that the sediment particles start to move when the bed shear stress becomes
larger than a critical value which depends on the parameter $\sqrt{ \left( \rho^*_s/\rho^*-1\right) g^* d^{*3}}/\nu^*$, which is known as both Galilei/Galileo
number and sediment Reynolds number. Hence,
the shear stress just above the small spheres is evaluated as a function of time and figure \ref{fig7} shows its streamwise component at different phases 
of the cycle for $D=28, d=2.6$ and $R_\delta=56$. 
As expected, the numerical results show that the shear stress acting on the seafloor (far from the large sphere) attains its 
maximum value with a phase shift $\phi \simeq \pi/4$ with respect to the free stream velocity. Indeed, the flow regime in the bottom boundary 
layer keeps laminar for $R_\delta=56$ and even for $R_\delta=112$. The spatial and temporal distribution of the shear stress
feels the effects of the large sphere and the action of the vortices shed by the sphere itself. Indeed, the largest values of the shear stress are found 
around the sphere in the lateral regions where the velocity attains its maximum values, while behind the sphere, regions of relatively small 
values of the shear stress are observed because of the shadow effect of the sphere which creates a sheltered area characterized by a relatively small velocity.
Moreover, the footprints of the coherent vortex structures shed by the sphere can be easily identified, since the coherent vortices induce locally relatively 
high velocities and hence high values of the shear stress.  

For example, at $t=11.98$, the maximum values of the bed shear stress are attained in the vicinity of the large sphere and into two
layers, which leave the lateral surface of the sphere in the downstream direction, because of the local large velocities induced by the sphere
presence and the flow separation. However, at $t= 10.21$, when the shear stress far from 
the sphere is relatively small because of the flow inversion, the regions characterized by the largest values of the shear stress are in the wake of the 
sphere and can be paired with the coherent vortices convected by the free stream. 

\begin{figure}
\begin{picture}(0,210)(0,0)
\ifnine
\putpic{-1,108}{.52}{figure18a}
\put(-10,160){$x_2$}
\putpic{197,108}{.52}{figure18b}
\putpic{-1,5}{.52}{figure18c}
\put(-10,58){$x_2$}
\putpic{197,5}{.52}{figure18d}
\put(91,-2){$x_1$}
\put(289,-2){$x_1$}
\fi
\end{picture}
\caption{Shear stress acting on the plane $x_3=3.08$, just above the crest of the small roughness elements, shadowed by colours for $D=28, d=2.6, R_\delta=56$. From left to right and from top to bottom: $t=9.82$, $t=10.21$, $t=11.39$, $t=11.98$.}
\label{fig7}
\end{figure}
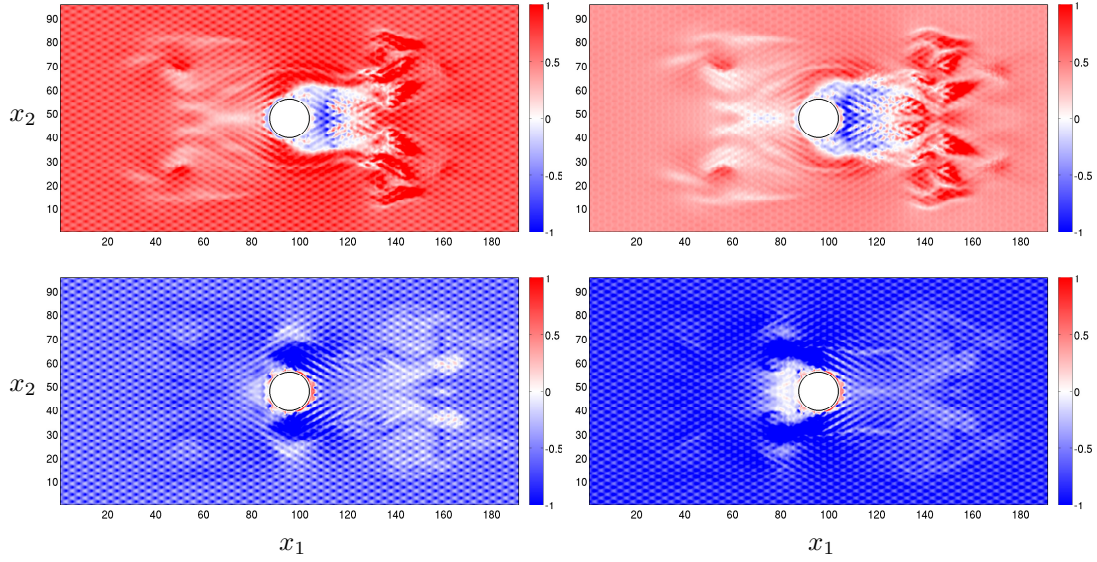
\begin{figure}
\begin{picture}(0,260)(0,0)
\ifsixteen
\putpic{30,94}{.8}{figure19a}
\putpic{30,3}{.8}{figure19b}
\put(185,-4){$x_1$}
\put(16,178){$x_2$}
\put(16,50){$x_2$}
\put(25,250){$a)$}
\put(25,85){$b)$}
\fi
\end{picture}
\caption{Top view of the bottom areas where the small spheres could be potentially mobilized by the flow during certain phases of the 
oscillatory period, in which the Shields parameter $\theta_{max}$ exceeds $\theta_{crit}=0.05$, for either the relative density of the 
spheres $s=1.025$ (grey areas) or $s=1.05$ (red/darker areas) and $D=28, d=2.6$. 
The broken line indicates the projection of the large sphere equator $a)$ $R_\delta=56$ and $b)$ $R_\delta=112$.
}
\label{fig14}
\end{figure}
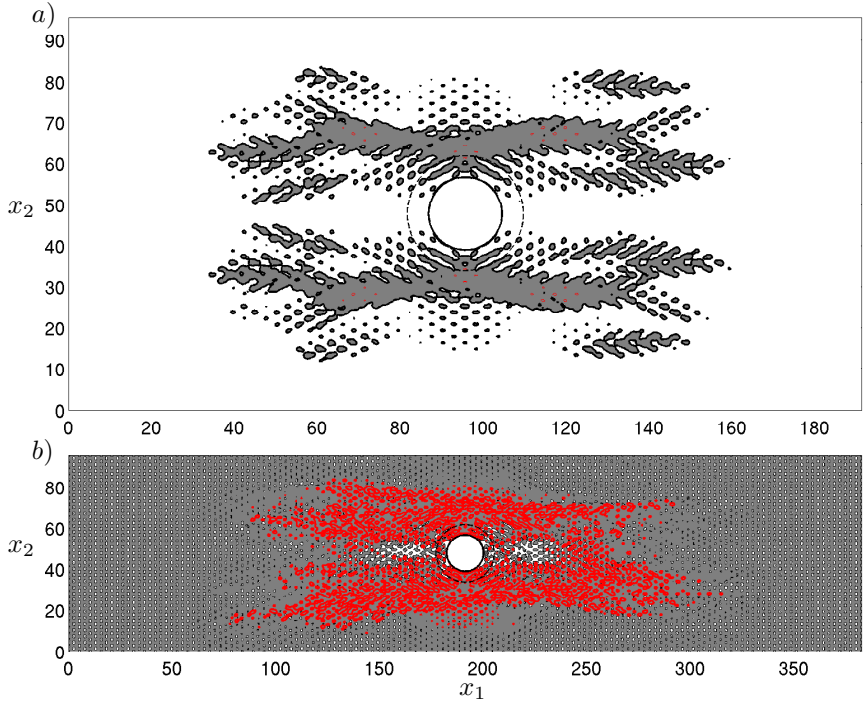

Finally, the region where the flow would be able to
move the small spheres at any phase of the cycle is visualized in Figure \ref{fig14}a. This region is evaluated by computing the modulus of the shear stress at
each phase and selecting its maximum value $\tau^*_{max}$, during the oscillatory cycle, at each location. Then, the maximum value of the Shields parameter 
$\theta_{max}$ is computed as $\theta_{max}=\tau^*_{max}/(\rho^*_s-\rho^*)g^*d^*$ and the region of the possible motion of the small spheres 
is assumed to be coincident with the area of the sea bottom where $\theta_{max}$ is larger than a critical value $\theta_{crit}$. The value of
$\theta_{crit}$ depends on the particle Reynolds number $R_p=\sqrt{(\rho^*_s/\rho^*-1)g^*d^{*3}}/\nu^*$
and different empirical relationships are available into the literature to determine $\theta_{crit}$.
One of the most used is that of \citet{brownlie1981} which has been recently amended by \citet{parker2003} dividing 
its value by a factor $2$. For small values of $R_p$, the relationships provide large values of $\theta_{crit}$. However,
\citet{soulsby1997} noticed that, even for small grain sizes, the value of $\theta_{crit}$ never exceeds $0.3$. To take 
into account this experimental evidence, he proposed a new relationship. 
Because of the large uncertainty which affects the estimate of $\theta_{crit}$, the results of Figure \ref{fig14} 
are obtained by using $\theta_{crit}=0.05$ which is a reasonable estimate for relatively large spherical particles.

Because of the relatively large size of the small spheres and the weakness of the oscillatory flow, the region of potential erosion around the large sphere
is quantified by assuming that the small spheres are made by a relatively light material (e.g. polyethylene with a relative density $s=\rho^*_s/\rho^*$ equal 
to both $1.025$ and $1.05$). 

As it could be guessed on the basis of the time development of the streamwise component of the bottom shear stress (see figure \ref{fig7}), the results
plotted in figure \ref{fig14}a show that the small spheres, simulating the sediments around the "large" spherical object, tend to be swept away mainly from the sides of the
large sphere and along the trajectories of the coherent vortex structures shed by the object. The symmetry of the flow is not perfect because it would be necessary 
to simulate more cycles to have a fully periodic regime.
For $R_\delta=56$, the small spheres, which simulate the sediment grains, would move only for $s=1.025$. Moreover, sediment transport would be observed only 
in the region surrounding the large sphere while the sediment grains far from it would be at rest (clear water conditions).
On the other hand for $R_\delta=112$ and $s=1.025$ (see figure \ref{fig14}b), the small spheres would move also far from the large sphere 
even though the largest values of the shear stress and the most
intense sediment transport would be observed close to the large sphere. However, increasing $s$ ($s=1.05$), the sediment motion would again be confined in the
region close to the large sphere and clear water conditions would be observed.

Finally, we should point out that the critical value of the Shields parameter is evaluated on the basis of an empirical approach
which is based on data obtained for steady flows and does not take into account the effects of the pressure gradient oscillations \citep{sleath1999}.
Even though it is likely that the oscillatory character of the flow has no significant influence on the initiation of sediment motion because
of the large ratio between the fluid displacement oscillations and the sediment size, future simulations with mobile particles might provide detail information
on the conditions of incipient motion of sediment particles in oscillatory flows, which is still an active area of research \citep{frank2015,frank2015b}.

To better simulate the flow close to and within the sea bottom, the run for $D=28, d=2.6, R_\delta=112$ was 
repeated with five layers of sediment particles, laid over the plane rigid wall located at $x_3=0$.
If the results obtained for these values of the parameters are compared with those previously discussed, no qualitative change is observed. 
In particular, the boundary layer separates from the surface of the large sphere and generates free shear layers which in turn roll up and generate large
vortex structures. Later, these vortex structures break and give rise to turbulent regions. These turbulent regions are advected by the mean flow
until viscous effects damp turbulence oscillations. 
It is only interesting to point out that the streamwise pressure gradient, which forces the flow,
is independent of the vertical coordinate and it tends to generate fluid motion also inside the packed small spheres. 
However, the fluid encounters a large resistance through the sphere gaps and the flow through the 
particles turns out to be negligible.

\begin{figure}
\begin{picture}(0,510)(0,0)
\ifeighteen
\put(0,-55){
\put(0,130){
\put(50,330){
\put(0,5){\includegraphics[trim=0cm 0cm 13cm 3cm,clip,width=.7\textwidth]{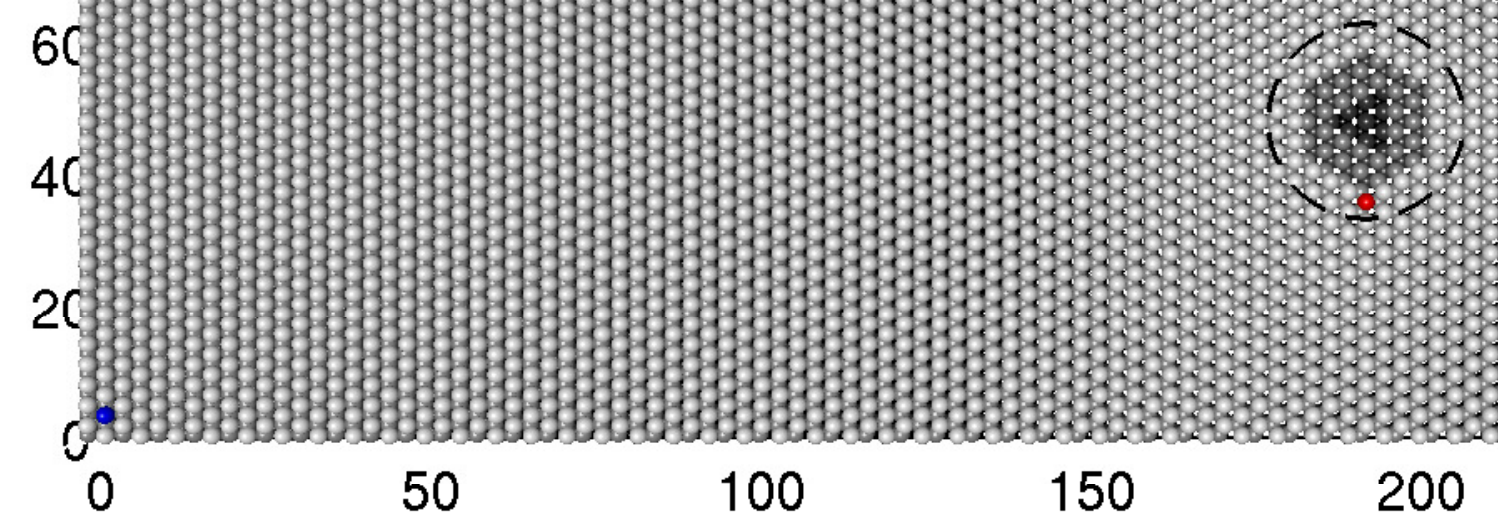}}
\put(-8,90){$a)$}
\put(136,-2){$x_1$}
\put(-5,54){\rotatebox{90}{$x_2$}}
\put(243,44){\mut{\vector(0,1){12}}}
\put(19,42){\mmt{\vector(0,-1){12}}}
}
\putpic{75,216}{.585}{figure20b}
\putpic{70,110}{.6}{figure20c}
\put(60,308){$b)$}
\put(60,204){$c)$}
\put(55,248){\rotatebox{90}{$\sqrt{F^2_1+F^2_2}$}}
\put(55,164){\rotatebox{90}{$F_3$}}
}
\putpic{70,135}{.6}{figure20d}
\put(60,228){$d)$}
\put(55,185){\rotatebox{90}{$T_2$}}
\put(60,121){$e)$}
\putpic{70,59}{.6}{figure20e}
\put(187,50){$t$}
\put(55,93){\rotatebox{90}{$U_e$}}
}
\fi
\end{picture}
\caption{
Top view of the lower-left part of the bottom (panel $a$). Red (on the right) and blue (on the left) spheres are numerically instrumented to
determine the hydrodynamic force exerted by the flow oscillations. The broken black line indicates the equator of the large sphere
($R_\delta=112, D=28, d=2.6$ and $5$ layers of small spheres arranged in a hexagonal patterns).
Time development of the horizontal ($\sqrt{F_1^2+F_2^2}$) (panel $b$) and vertical ($F_3$) (panel $c$) components of the hydrodynamic 
force as well as the spanwise torque component ($T_2$) (panel $d$) acting on the small spheres indicated in panel $(a)$ (blue-crosses and red-circles lines are related to the blue (left) and red (right) spheres, respectively) 
located either in the top layer $[$\textbf{---}$]$ or in the bottom layer $[$- - -$]$. 
Panel $e$ shows the velocity far from the bottom.
}
\label{figYY}
\end{figure}
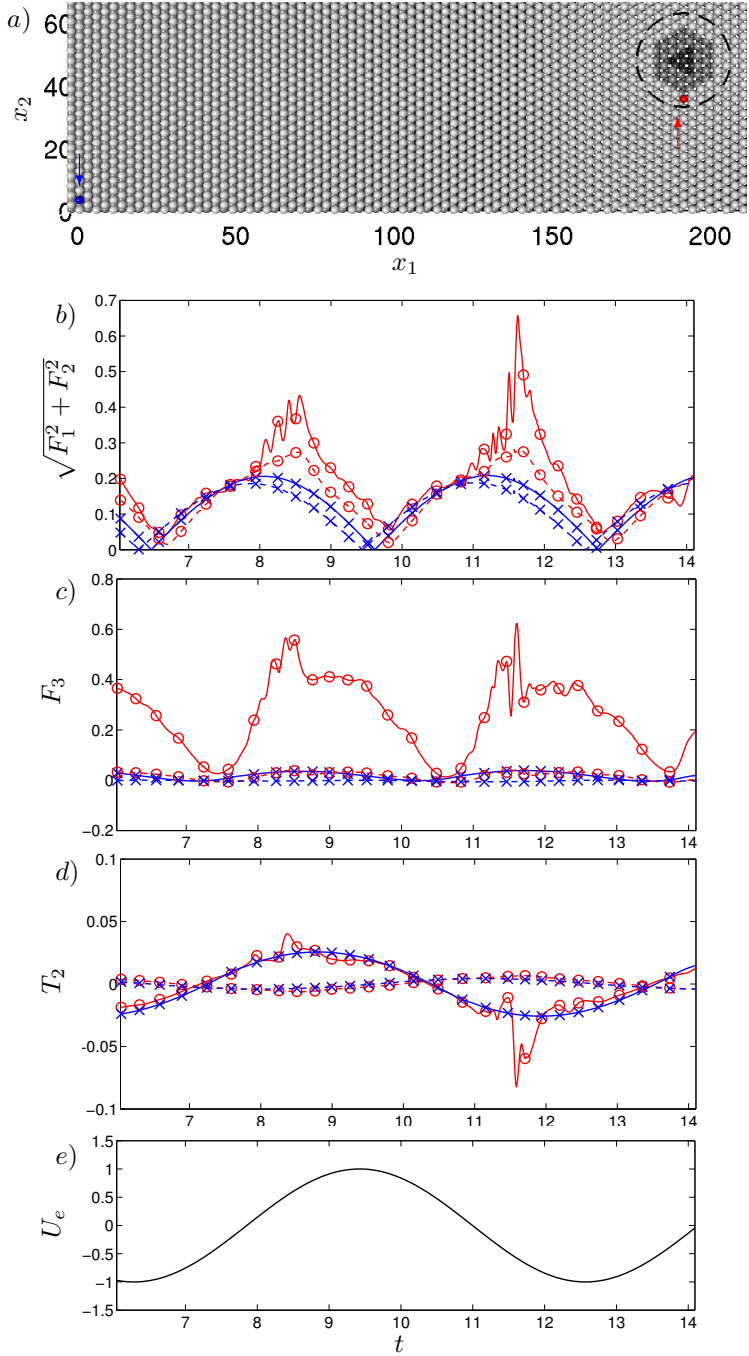

It is also interesting to analyse the force acting on a single small sphere within the bed.
Figure \ref{figYY} shows the time development of the dimensionless horizontal component ($\sqrt{F_1^2+F_2^2}$) of the force (panel $b$) along with
the vertical ($F_3$) component (panel $c$) 
and the spanwise component of the torque acting on the sediment grains considered in panel $a$ of figure \ref{figYY}. 
The force $F_i$ and torque $T_i$ components acting on the large and small spheres are evaluated by means of the numerical integration of $t_i$
and $\delta_{ijk} x_j t_k$ on the surface of the spheres:
\[\displaystyle
F_i=\frac{8 F^*_i}{\rho^* \left(U^*_0\right)^2 \pi d^{*2}} = \frac{8}{\pi d^2}\int_S t_i d S\:, \ \ \ \ \ \ \ T_i = \frac{16 T^*_i}{\rho^* \left(U^*_0\right)^2 \pi d^{*3}} = \frac{16}{\pi d^3}\int_S \delta_{ijk} x_j t_k d S\:,
\]
where $t_i= -p n_i + \frac{1}{R_\delta} \left( \frac{\partial u_i}{\partial x_j} + \frac{\partial u_j}{\partial x_i}\right)n_j$ is the sum
of two contributions. The former is due to the pressure acting on the surface of the spheres and the latter to the viscous stresses. 
In the previous relationships $S$ indicates the surface of the spheres, $n_i$ is the component along the $x_i$-axis of the unit vector normal to the surface,
$x_i$ indicates the position of the infinitesimal surface of the sphere with respect to its centre and $\delta_{ijk}$ is the Levi-Civita symbol. 
The blue lines (with crosses) show the time development of the components of the force and torque acting on the particles
far from the large sphere (blue particles on the left hand side of panel $a$) while the red lines (with circles) show the force components on the particles close to the large sphere
(red particles on the right hand side of panel $a$). In both panels $b$ and $c$, continuous lines and broken lines appear. The former refer to particles in the top
layer while the latter refer to particles in the bottom layer. Far from the large sphere, the horizontal component of the force is almost
in phase with the pressure gradient which originates the fluid motion (panel $e$ of figure \ref{figYY} shows the free stream
velocity). The differences between the horizontal component of the force
acting on the particles of the surficial layer and that acting on the particles of the bottom layer are small, thus indicating that
the viscous contribution to the force is small. Moreover, the vertical component is negligible. Close to the large sphere, the force
acting on the particles shows large fluctuations which are due to the interaction of the particles with the vortex structures shed by the large
sphere. Of course the largest fluctuations are observed for the particle of the surficial layer, since the effects of the vortex structures on the
particle rapidly damp moving inside the packed particles.
We should note the presence of large uplifting forces near $t=8.5$ and $t=11.5$ that can significantly reduce the particle stability.
The time development of the force acting on the surficial particle close to the large sphere indicates a phase lag of the maximum force 
with respect to the maximum flow which is similar to that observed
by \citet{Mazzuoli2016} for larger values of the Reynolds number. Indeed, the presence of the large sphere causes a local acceleration of the 
flow around it and the small particles feel larger values of the velocity which induce also significant values of the vertical component of 
the force.
The torque oscillates and is significant 
only for the particles in the surficial layer. In particular the torque oscillates almost in phase with the free stream velocity even though, close
to the large spherical object, large fluctuations are present which are caused by the effects of the shear layers and the vortex structures shed by the 
large sphere.

\begin{figure}
\begin{picture}(0,260)(0,0)
\iftwenty
\put(0,150){
\put(30,10){\includegraphics[trim=0cm .5cm 0cm 1cm,clip,width=.93\textwidth]{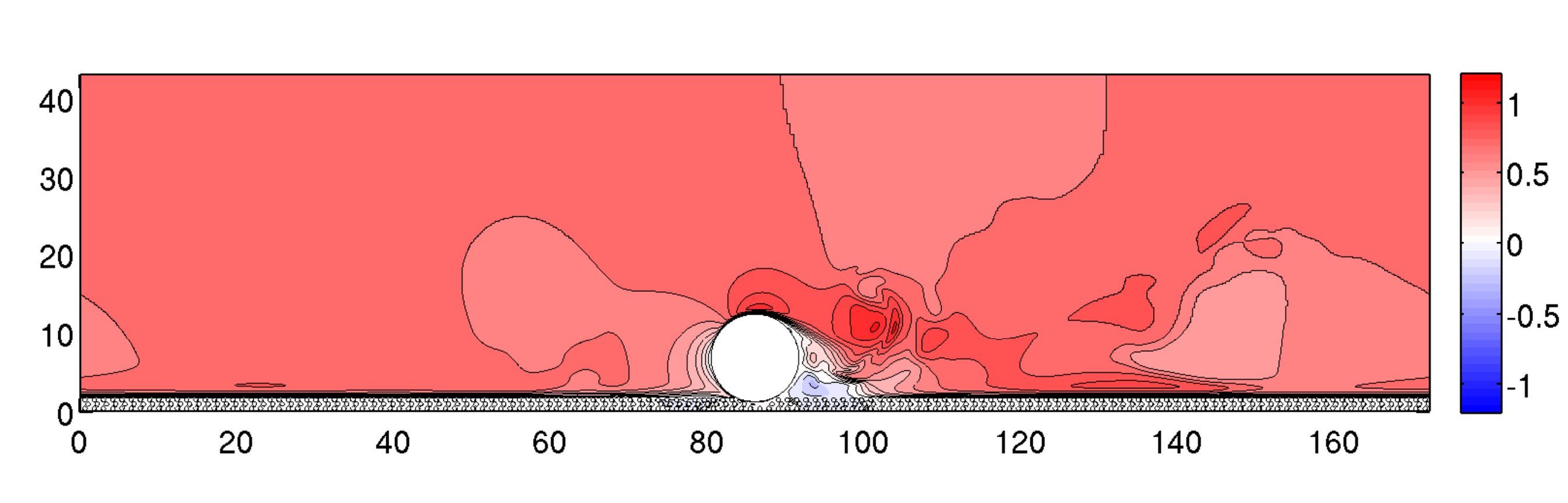}}
\put(25,95){$a)$}
\put(25,62){$x_3$}
}
\putpic{40,0}{.8}{figure21b}
\put(30,138){$b)$}
\put(30,74){$\boldsymbol{F}$}
\put(198,-4){$t$}
\put(198,155){$x_1$}
\fi
\end{picture}
\caption{Panel $a)$ shows the streamwise velocity component at $t=2.75\pi$ in the middle vertical plane crossing the largest sphere for $R_\delta=112$, $D=11.21$, $d=0.56$, $\epsilon = 0.112$. The thin lines are the iso-velocity contours equispaced by $\Delta u_1=0.1$. Panel $b)$ shows, for the same simulation, the streamwise (thick solid line), spanwise (broken line) and wall-normal (thin solid line) force components acting on the large sphere throughout the last period presently simulated. The grey broken line indicate the free-stream velocity.}
\label{fig18}
\end{figure}

A simple dimensional analysis of the phenomenon shows that the flow field depends on three dimensionless parameters
beside the geometrical arrangement of both the large sphere and the small spheres. 
In particular, the results would appear to depend on the value of $\epsilon$, i.e. the dimensionless distance of the spheres from the bottom and
of a sphere from the other. However, as discussed previously, the parameter $\epsilon$ has a minor influence on the velocity field and the forces acting on both
the large sphere and the spherical roughness elements. Hence, in the following, we do not consider explicitly the values of $\epsilon$ which
are slightly different in the simulations described in the paper.
We chose the Reynolds number $R_\delta=U^*_0\delta^*/\nu^*$
and the ratios $D=D^*/\delta^*, d=d^*/\delta^*$ as dimensionless parameters and we fixed the sphere arrangement as depicted in figure \ref{fig1}.
The reader should be aware that the Keulegan-Carpenter number $K_c=U_0^*/(\omega^*D^*)$ of the phenomenon, often introduced
in coastal engineering studies, turns out to be $R_\delta/(2D)$ and the Reynolds number $Re=U_0^{*2}/(\omega^* \nu^*)$ turns out to be
equal to $R_\delta^2/2$.
The large computational costs did not allow an exhaustive investigation of the parameter space to be carried out. 
Hence, being interested in the flow field generated close to the sea bottom by wind waves and in its interaction with small spherical objects,
the simulations we carried out consider values of $R_\delta$ typical of the boundary layer under sea waves. Moreover,
values of $K_c$ of order $1$ are considered such that the velocity and vorticity fields are significantly affected by
the unsteadiness of the forcing flow and the wake behind the sphere is topologically different from that which is generated behind an isolated sphere in a steady
uniform flow.
To show that the numerical approach can be used to investigate cases of practical relevance, a further run was carried out
by considering $D=11.21, d=0.561$, $R_\delta=112$ and $\epsilon = 0.112$. These values of the parameters correspond to sea waves characterized by a period and a height equal
to about $10$ s and $0.4$ m, respectively, propagating on a water depth $h^*\simeq 30$ m. 
Moreover, the sediment diameter $d^*= 1$ mm is that of a coarse sand and $D^*=20$ mm corresponds to the size of
a small munition.
Figure \ref{fig18}a shows a snapshot of the velocity field around the small munition at $t=2.75 \pi$. Looking at the velocity field, 
it is possible to understand why the streamwise component $F_1$ of the force on the munition (see figure \ref{fig18}b) attains relative
maxima for values of $t$ close to $n \pi$ ($n=1,2, ..$), when a region of dead water develops behind the munition. Moreover, it appears that the small bumps 
and further relative maxima in the time development of $F_1$ are caused slightly before other maxima by the passage of clockwise/counter-clockwise vortices 
shed during the previous half cycle and dragged by the free stream. 
The lift force is always positive and tends to pick-up the small munition, while the spanwise component of the force
almost vanishes even though small oscillations are randomly generated when the symmetry of the flow, with respect to a vertical plane crossing the
centre of the large sphere, is broken by the instability of the wake.
A detailed analysis of the flow field and the bed shear stress shows that no
qualitative change, with respect to the results previously described, is generated by the variation of the parameters, at least
in the investigated range.

\section{Conclusions}

The numerical results previously described show that the direct numerical simulation of the oscillatory flow around
an object lying on small spherical particles, which mimic a sandy bottom, can be performed thus making possible a detailed 
investigation of this unsteady, three-dimensional flow field. The power of actual computers allows
only the simulation of relatively small objects, coarse sand and moderate values of the Reynolds numbers. 
Nevertheless, problems of practical relevance can be tackled and solved.

The oscillatory flow around the object gives rise to coherent vortex structures which are generated by the roll up
of the free shear layers shed by the object surface because of boundary layer separation. These vortices move away 
from the object and later break-up and generate turbulence.
Even though the Reynolds number is not large enough to lead to a turbulent flow within the bottom boundary layer,
turbulence is observed also close to the bottom but only below the large vortices shed by the object. 
Accurate information on the possible incipient motion of the sediment particles are provided by the analysis of the spatial
and temporal distribution  of the bottom shear stress and of the force and torque acting on the particles. 
The area of the possible erosion around the large spherical object resting on the bottom depends on the parameters
which characterize the flow, the object and the sediment particles. First of all, for the particles to be
set into motion, the maximum value of the Shields parameter should be larger than its critical value for the 
initiation of sediment transport. Because of the presence of the large, resting object, the flow is accelerated
around the object and therein the Shields parameter is significantly larger than far from it.
Hence, close to the large sphere, the sediment can be set into motion even though far from the sphere the flow is not strong enough
to move the sediment (clear water condition). Even though only moderate values of the flow Reynolds number are simulated, results
for both clear water conditions and live bed conditions (sediment is set into motion also far from the object) are obtained
by varying the relative density of the sediment particles and considering hydrodynamic and morphodynamic parameters which can be
easily reproduced in a laboratory experiment. Since, the erosion and deposition processes are controlled by the divergence of the
sediment transport, the area of possible erosion is limited also in the case of live bed conditions.
The Keulegan-Carpenter number, $K_c$, largely controls the area of possible erosion 
since 
the displacement of the vortex structures shed by the large sphere increases as the 
value of $K_c$ is increased.
Moreover, the Keulegan-Carpenter number
affects also the trajectories of the vortices which do not follow the path of the previous cycle
if the flow Reynolds number is large enough to generate a chaotic flow and to trigger transition to turbulence. The 
numerical simulations show that the drag
as well as the lift forces, acting on the sediment particles close to the large sphere, are characterized by large fluctuations 
and vary from particle to particle, thus showing the possible selective pick-up of the sediments by the forcing flow.
Finally, it is worth pointing out that a significant reduction of both the drag and lift forces is observed moving within the
sea bottom from the top layer of particles towards the underlying layer because of the sheltering effects of the particles which stand above.
The next step of the research project is to let the particles to move and to investigate their dynamics and the time development of
the scour around the object.\\

This study has been funded by the Office of Naval Research (U.S.A.) under the research project n. N62909-14-1-N126.
Moreover, J. Simeonov and J. Calantoni were supported, under base funding to the Naval Research Laboratory, by the Office of Naval Research and M. Mazzuoli and P. Blondeaux, under the project PRA2014, by the University of Genoa 

\bibliographystyle{jfm}
\bibliography{\bibdir Refbib}

\end{document}